\documentclass{aa}  
\usepackage{graphicx}
\usepackage{txfonts}
\usepackage[colorlinks=true,linkcolor=blue,citecolor=blue,urlcolor=blue]{hyperref}
\usepackage{soul}
\usepackage[normalem]{ulem}

\usepackage{color}
\usepackage{amstext}

\begin{document} 

  \title{Hot molecular hydrogen in the central parsec of the Galaxy through near-infrared 3D fitting}
  \author{A. Ciurlo, T. Paumard, D. Rouan \and Y. Cl\'enet
          }
 \institute{LESIA, Observatoire de Paris, PSL Research University, CNRS, Sorbonne Universit\'es, Univ. Paris Diderot, UPMC Univ. Paris 06, Sorbonne Paris Cit\'e, 5 place Jules Janssen, 92195 Meudon, France\\
              \email{anna.ciurlo@obspm.fr; thibaut.paumard@obspm.fr; daniel.rouan@obspm.fr}
             } 
  \date{}

 \titlerunning{Molecular gas in the central parsec}

  \abstract
   {}
   {We have investigated neutral gas in the central cavity of the circumnuclear disk (CND) at the Galactic Center, where the ionized minispiral lies, to describe the H$_{2}$ distribution and properties in this ionized environment.}
   {This study was carried out through a spectro-imaging data cube of the central cavity obtained with SPIFFI on the VLT. The observed field of view is $36\arcsec\times29\arcsec$, with a spectral resolution R~=~1\,300 in the near-infrared. These observations cover several H$_2$ lines. To preserve the spatial resolution and avoid edge effects, we applied a new line-fitting method that consists of a regularized 3D fitting. We also applied a more classical 1D fitting to compare the relative strength of the H$_{2}$ lines.}
   {We present high spatial and spectral resolution maps of the intensity, velocity, and width of five H$_{2}$ lines and an extinction map derived from H$_{2}$. Molecular gas is detected everywhere in the field. In particular, in addition to the known CND features, we detected an emission from the northern arm cloud and from the minicavity. The excitation diagrams allow us to estimate the temperature, mass, and density of these features. }
 {We interpret the CND emission as coming from a hot, thermalized, thin layer at the surface of the clouds. The observed H$_{2}$ corresponds only to a small fraction of the total H$_{2}$ mass. The emission remains fairly strong in the whole central cavity, but it is not thermalized. A strong deviation from thermal equilibrium is detected near the minicavity. We suggest that this emission is caused by constantly forming H$_{2}$ that is destroyed again before it reaches ortho/para equilibrium.}


   \maketitle


\section{Introduction}
\label{intro}

In the central few parsecs of the Galaxy the gas is organized into two main structures: the circumnuclear disk (CND) and Sgr~A~West (the minispiral). 

The CND is an asymmetric and very clumpy gas ring on a scale of 2~--~5~pc, composed of molecular clouds (\citealt{Gatley86}, \citealt{Guesten87}, \citealt{Yusef-Zadeh01}). 
At first approximation, it is in circular rotation around the radio source Srg~A* at about 100~km s$^{-1}$ \citep{Jackson93}. 
The CND inner cavity -- also referred to as the central cavity -- is ionized and encloses the minispiral. 

Sgr~A West is composed of clouds of atomic and ionized gas \citep{Lo83}. 
It consists of the ionized surface  layers of larger neutral clouds that are drawn by tidal force toward the Galactic Center \citep{Paumard04}. 
The spiral aspect is caused by the projection in which we see these surface layers on the plane of the sky. 
More precisely, the northern arm of the minispiral is probably located at the surface of a cloud that is falling toward the center and is exposed to intense UV radiation. 

The central cavity is known to be embedded in a strong radiation field because a central cluster of young stars provides ionization for both the CND inner cavity and the minispiral \citep{Martins07}.
The CND and the minispiral are strongly connected \citep{Ekers83}: the western arc of Sgr~A~West is an ionized layer at the surface of the inner edge of the CND \citep{Guesten87}. 
The CND is believed to be fueled by infall from dense molecular clouds 10 pc away. 
The CND, in turn, releases material into the central cavity. 
Here, illuminated by the UV field, it is at the origin of the observed minispiral \citep{Guesten89}. 
This suggests that the two structures have common or related origins.

We have investigated molecular hydrogen, H$_2$, in the ionized central cavity that is delineated by the CND, where the minispiral lies. 
In this environment the strong ultraviolet (UV) radiation is expected to destroy H$_{2}$.  
However, in principle, hydrogen molecules could be found within the neutral clouds whose ionized borders delineate the arms of the minispiral. 
Here they could be shielded from the radiation by an ionization front and thus survive dissociation. 
H$_2$ could also be found in the transition zone between the CND and the minispiral or on the line of sight.

H$_2$ emission has previously been detected inside the central cavity by \cite{Gatley86}. 
 \cite{Jackson93} analyzed HCN and \ion{O}{i} lines, which led to the detection of a large amount of neutral (300~M$_{\odot}$) gas associated with the streamers of the minispiral. 
It also showed several dense and distinct streamers at the edge of the CND. 
\cite{Marr92} revealed molecules in the bar of the minispiral
that are shielded by dust in the ionized features. 
\cite{Yusef-Zadeh01} clearly detected H$_2$ in the central cavity and showed that its kinematics is inconsistent with the CND direction of rotation.

Our goal is to extend the search and study of H$_2$ in the central parsec by characterizing the gas distribution, dynamics, and excitation. 
This study is carried out by analyzing a SPIFFI \citep{Tecza00,Eisenhauer03} spectroimaging dataset in the H and K bands that covers several H$_{2}$ lines. 

In Sect.~\ref{obs} we present these observations together with the calibration we applied. 
The analysis was made in two steps. 
In the first part (Sect.~\ref{morph}), the gas morphology and dynamics were analyzed. 
We applied an original line-fitting method to handle the noise and conserve good angular resolution: a global regularized 3D
fitting (\citealt{Paumard14}, 2016 in prep.). 
We eventually obtained high-resolution maps of the gas flux, velocity, and line width. 
The comparison of two H$_{2}$ lines (1--0~S(1) and Q(3)) allowed us to correct the extinction effect.
In the second part of the analysis (Sect.~\ref{prop}), we studied the possible excitation mechanism through the simultaneous analysis of eight H$_{2}$ lines to which we applied a more classical 1D spectroscopic model fitting. 
The derived fluxes allow tracing the excitation diagrams of several areas on the field of view and computing the relative excitation temperature and mass. 
We conclude on molecular gas morphology and the variety of emission processes in Sect.~\ref{concl}.
 
\section{Observations and dataset reduction}
\label{obs}

\subsection{Dataset}

The dataset consists of a near-infrared (NIR) spectroimaging cube with two spatial dimensions and a wavelength dimension. 
It has been obtained with the Spectrometer for Infrared Faint Imaging (SPIFFI), a NIR integral field spectrograph installed on the VLT \citep{Tecza00}. 
This cube was acquired during the first observing runs as a guest instrument at the VLT in March and April 2003. 
After these observations SPIFFI was equipped with adaptive optics and became SINFONI. 
The same dataset has already been used for another program about stellar populations \citep[among others]{Eisenhauer03,Horrobin04,Paumard06}. 

The central parsec has been observed for two nights for one hour each, resulting in a mosaic of $\sim35\arcsec\times35\arcsec$. 
The southwestern (from 20$\arcsec$ east to 15$\arcsec$ south of Sgr~A*) and northwestern (from 15$\arcsec$ north to 15$\arcsec$ west of Sgr~A*) corners as well as the west side (from 12$\arcsec$ west of Sgr~A*) of the SPIFFI field of view cover part of the CND (as can be inferred by comparing to \citealt{Liszt03, Christopher05}). 
The rest is occupied by the central cavity.
The observations are seeing-limited (FWHM 0.75\arcsec) with a 0.25\arcsec~pixel sampling and a spectral resolution R~=~1\,300 in the combined H+K mode.
        
This spectral range includes many H$_2$ lines. 
The lines we were able to detect in our dataset are reported in Table \ref{tab:lines}.

                \begin{table}[htdp]
                \begin{center}
                \begin{tabular}{|c|c|c|c|}
                \hline 
                Transition & Wavelength [$\mu$m] & avg S/N & max S/N \\[6pt]
                \hline
                1--0~S(3)        &       1.9575  &       2.6     &       9.1     \\[2pt]
                1--0~S(2)        &       2.0337 &        1.1 &   6.4     \\[2pt]
                1--0~S(1)        &       2.1217 &        5.3 &   30.8    \\[2pt]
                1--0~S(0)        &       2.2232 &        1.9 &   8.0     \\[2pt]
                2--1 S(1)        &       2.2476 &        1.0 &   3.2     \\[2pt]
                1--0~Q(1)        &       2.4065 &        3.9 &   23.9    \\[2pt]
                1--0~Q(2)        &       2.4133 &        0.9 &   4.9     \\[2pt]
                1--0~Q(3)        &       2.4236 &        4.4 &   22.3    \\[2pt]
                \hline
                \end{tabular}
                \end{center}
                \caption{Transition name, wavelength, and average and maximum signal-to-noise (S/N) over the field for H$_{2}$ lines covered by the SPIFFI spectral range.}
                \label{tab:lines}
                \end{table}%

The H$_{2}$ molecule has two spin states: the ortho, triplet, with parallel nuclear spins, and the para, singlet, with antiparallel nuclear spins. 
Odd-numbered transition lines correspond to ortho states, even-numbered match para states.
In our dataset, most para lines and 2--1 S(1) have a low signal-to-noise ratio (S/N), which makes the analysis of these lines possible only in the zones of highest intensity. 

The spectrum also contains recombination lines such as Br$\gamma$, \ion{Fe}{iii,} and \ion{Fe}{ii}, which will be  the object of another study.

The observations were processed using the SPIFFI pipeline \citep{Modigliani07}; the reduced data-cube is the same as in \cite{Eisenhauer03}.

\subsection{Calibration}

To compare different lines, the flux has to be dereddened and calibrated in photometry.
The data that we obtained were already reduced and partially calibrated to remove the atmospheric absorption feature between the H and K bands. 

We have completed the calibration using stars in the field chosen from the catalogs of \cite{Paumard06}, \cite{Blum03}, and \cite{Bartko09}. 
The calibration was done in two steps: the first step was to correct the slope of all spectra (relative calibration, Sect.~\ref{relcal}), the second step was to convert the analog-to-digital unit into physical flux units (photometric calibration, Sect.~\ref{cal}).
        
        \subsubsection{Relative calibration} 
        \label{relcal}
        
        To carry out the relative calibration we divided the datacube by the observed spectrum of a standard spectrometric calibrator and multiplied it by the intrinsic -- or expected -- spectrum of this calibrator.
        
        At this stage, we only considered the shape of the spectra. The absolute flux calibration comes at a later stage. To measure the observed spectrum of each calibrator, we integrated over a 
        diamond-shaped aperture of 5 pixel 
        width and height 
        (13 spatial pixels). For background estimation and subtraction, we used a 5-pixel-wide square aperture minus the 
        diamond-shaped aperture used to estimate the stellar spectrum itself.
        
        The intrinsic spectral shape of each star was estimated using a blackbody curve, at the effective temperature of the star. 
        For high temperatures, the H- and K-band emission is in the Rayleigh-Jeans tail of the blackbody, where the shape of the curve does not change with temperature. 
        Therefore, we considered the same temperature (35\,000~K) for every calibration star. 
        This blackbody was reddened using a power-law extinction curve for the NIR: A$_{\lambda}\sim\lambda^{-\alpha}$, where $\alpha\sim2.07$ \citep{Fritz11}. 
        The local extinction is different for each star and  was taken from the extinction map in \citet{Schodel10}.
        
        \subsubsection{Photometric calibration}
        \label{cal}
        
        From the resulting data cube we extracted the spectrum of cool stars of known magnitude for the absolute calibration. 
        We considered the same diamond-shaped aperture as before. The measured flux, integrated over the K band, was compared to the expected one (computed from the known star magnitude). 
        
        This comparison was made taking into account that, for our aperture, the enclosed energy is estimated to be $\sim$~25\% of the total. 
        To calculate this factor, stars of the field are not well-suited since they are not isolated. 
        Therefore we created a synthetic point spread function
(PSF) of the same width as the stars on the field (1.5~$\pm$~0.2~pixels, computed from the fit of a 2D Gaussian over few stars). 
         The enclosed energy in the aperture was then evaluated from this synthetic PSF.
        
        \subsubsection{Final calibration factor and uncertainty}
        
        Three stars were used as relative calibrators:  GCIRS~16NW, GCIRS~16C, and GCIRS~16NE from \cite{Paumard06}. 
        Seven stars were used as photometric calibrators: stars 19, 28, 84, 91, 102, and 105 from \citealt{Blum03} and star 68 from \citealt{Bartko09}.
        
        We obtained a correction curve, meaning, a correction factor for each wavelength, for each combination of relative 
        and absolute calibration stars. 
        The average value of the eighteen curves was considered as the final relative correction.
        The uncertainty was evaluated as the dispersion of the curves of the different calibrators. It amounts to $\sim$~35\%. 
        This rather high value is dominated by the uncertainty on the stellar magnitudes and the exact stellar location within the aperture. 
        
        When considering the ratio of two spectral lines, the uncertainty due to the photometric calibration is no longer relevant because this error is common to all lines. 
        In this case, the uncertainty is only due to the dispersion of the slopes of the relative calibrators and amounts to $8\%$.

        To study the absolute flux and especially its spatial variations, the flux has to be corrected for extinction.
        To do so, we could consider values retrieved from the map of \cite{Schodel10} (or other available extinction maps), but this extinction map is obtained from stellar color excess. 
        The molecular gas is not necessarily situated at the same optical depth as these stars.      
        To properly correct the gradient of the extinction in the inner parsec, it is better to apply a method that is based on the dataset itself: considering the 1--0~Q(3) over 1--0~S(1) ratio. 
        This is detailed in Sect.~\ref{Schodel10}.
        
\section{Morphology and dynamics of the molecular hydrogen}
\label{morph}

Our analysis is based on a fit of each spectrum with a Gaussian function. In this way we obtain three parameters for each spatial
pixel: the  line flux, radial velocity, and width. 
We decided to search for line flux instead of the line intensity: line flux is more reliable than intensity because it does not depend on the instrumental resolution. 
After we obtained the three parameters for every pixel, we created for each parameter a map of its variations across the field.

\subsection{Method}

        \subsubsection{Regularized 3D fitting}
        
        The most common approach is to fit the spectrum of each spatial pixel individually. 
        The maps resulting from this approach are noisy.  
        Because the S/N in the SPIFFI dataset is not consistently high throughout the field of view, spatial smoothing has the effect of degrading the spatial resolution everywhere, increasing the correlation between neighboring pixels and introducing edge effects.
        
        To overcome these problems we applied a new method \citep{Paumard14}: a regularized 3D fitting. 
        We describe this method briefly, but more details will be given in Paumard et al. (2016, in prep.).
        
        The method consists of minimizing an estimator, $\varepsilon$, which is the sum of $\chi^2$ and a regularization term:         
        \begin{equation}
        \varepsilon(F,v,\sigma)= \sum_{\alpha, \delta, \lambda}\left[(D-M)\cdot W\right]^2+\sum_{a_{i}=F,v,\sigma} R_i(a_i)
        ,\end{equation}
        where the first term is the $\chi^2$ term (difference between the 3D dataset $D$ and the model $M$). 
        The elements of the cube are indicated by the spatial coordinates $\alpha$ and $\delta$ and by the wavelength coordinate $\lambda$. 
        $W$ is the weighting function that takes the S/N into account. $R_{i}(a_i)$ is the regularization term for the $a_{i}$-fitting parameter (flux $F$, velocity $v,$ and width $\sigma$). 

        The regularization is an L1--L2 algorithm borrowed from the deconvolution context, developed in \cite{Green90}, \cite{Brette96},
and \cite{Mugnier01} and generalized in \cite{Mugnier04} and coded in Yoda software by D. Gratadour\footnote{https://github.com/dgratadour/Yoda}. 
        The expression of this term for every parameter map is 
        \begin{equation}
        R_i(a_i) =  \mu_i \sum_{u=\alpha,\delta} \left[ \frac{\Delta O(u)_{a_i}}{\delta_i}-ln\left(1+\frac{\Delta O(u)_{a_i}}{\delta_i} \right) \right]
        .\end{equation}Here $\Delta O(u)_{a_i}$ is the spatial gradient of the $a_i$-parameter. 
        For weak gradients (small $\Delta O(u)$), the logarithmic term of L1--L2 dominates and the terms become a quadratic criterion, which penalizes gradients and smooths the map. 
        In contrast, when $\Delta O(u)$ is large, strong gradients are restored. 
        The transition from one regime to the other is adjusted through the $\delta_i$ and $\mu_i$ hyperparameters.
        
        This algorithm is very well suited to treat objects with strong and spatially correlated gradients such as sharp edges (e.g., planetary surfaces) or, in our case, ridges and streamers. 
        The algorithm disfavors random variations on the maps (strong gradients between pixels, which are noise) but still allows quick and coherent variations. 
        The fit over pixels with a low S/N is highly improved since the procedure uses neighboring points to adjust the solution. 
        This prevents overfitting of noise spikes. 
                
        However, there is one main inconvenience in this method: the necessary tuning of hyperparameters $\delta_i$ and $\mu_i$ for each parameter map (intensity, velocity, and width). 
        Six hyperparameters were therefore chosen. 
        They are not independent of one another, which complicates finding the solution. 

        $\mu_i$ regulates the smoothness of each parameter map. 
        An increase of $\mu_i$ has the immediate effect of smoothing the image, but it also has the secondary effect of limiting the variation range of the parameters. 
        Relative values of $1/\delta_i$ rescale each parameter causing them to vary around the same numerical order of magnitude. 
        A decrease of $1/\delta_i$ leads to stronger variations of the parameter across the field, but it also has the effect of lowering the weight of the regularizing term and to amplify the noise. 
        
        The challenge is to balance the six hyperparameters in order to gain sufficiently smooth maps while keeping fine spatial structures. 
        The aim is to obtain maps that appear smooth at a scale most possibly close to the spatial resolution.
        The procedure was run several times, with different combinations of hyperparameters, to determine the best strategy for identifying
a good hyperparameters combination. 

        The tuning of the hyperparameters is an iterative process.
        We describe below how we proceeded to determine good hyperparameters: firstly, we chose the initial guess as a set of constant maps. 
        The constant for each parameter is representative of the average value of each parameter over the area. 
        Secondly, we selected three values of  $1/\delta_i$ so that all three parameters varied in the same range. 
        The starting values of these hyperparameters was the order of magnitude of each parameter (for instance, 100 for a
line width that is expected to be around 120 km/s). 
        $1/\delta_i$ were then more finely tuned, trying to allow each map to cover as much of the selected range of variations
as possible. 
        When these ranges were well covered, the resulting maps were often noisy. 
        The next step was therefore to enhance the smoothing (through $\mu_i$) to avoid noise amplification. 
        The equilibrium between the six hyperparameters is delicate and needs several iterations.
       
        We currently do not have an objective and absolute criterion to select the best hyperparameters combination. 
        The reduced $\chi^{2}$ of the fit gives some indications, but is not enough to distinguish among distinct results that could have very close values of reduced $\chi^{2}$ (see, for instance, Fig.~\ref{res:zones}, where two models with close $\chi^{2}$ are compared to raw data).

        However, we have objective criteria that allow us to judge the result a posteriori, for instance, comparing the 3D model obtained with raw data. 
        Another strong objective a posteriori criterion to assess the method is given by comparing the 3D model with a classical 1D model. 
        Individual spectra in several zones can be extracted and fitted with a classical 1D model. 
        The regularized 3D fit model, averaged over the same zones, can then be compared to the classical 1D model, as shown in Sect.~\ref{3d1d}.
        One final, stronger a posteriori criterion is to compare the regularized 3D fitting results for distinct H$_{2}$ lines. 
        In Sect.~\ref{orthomaps} we show that 1--0~S(1), S(3), Q(1), Q(3), and S(0) all show the same large-scale structures in each parameter map. 

	Stars in the field might locally bias results by increased photon noise and because of some spectral features. Therefore the stars were masked. 
        The mask was built considering the root-mean-square (RMS) variation over the spectral direction at the location of each pixel (on spectral ranges free of features, especially stellar features). 
        Pixels that showed an RMS higher than a certain threshold were masked. 
        This threshold was chosen by comparing the resulting mask to the known position of stars (for instance, \citealt{Paumard06}) to cover the main bright stars of the region. 
        Our regularized 3D fitting allows interpolating over these masked regions.

                \begin{figure}
                \centering
                \includegraphics[width=9cm]{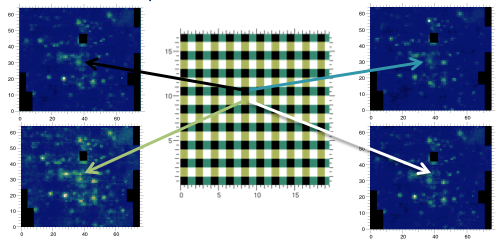}
                \caption{Four data cubes are created from dataset, taking one pixel out of four and interpolating over the holes. 
                The first cube comes from black pixels, the second from blue pixels, while the third and fourth cubes come from white and green pixels.}
                \label{res:4cubes}
                \end{figure}

        \subsubsection{Uncertainties}
        \label{EB}

        Since the method does not provide a direct estimate of the uncertainties, we developed a separate technique to evaluate them.
 
        We split the SPIFFI cube into four subcubes. 
        To do so, we selected only the odd columns and odd rows in the first cube, only the even columns and odd rows in the second cube, only the odd columns and even rows in the third cube and only the even columns and even rows in the fourth cube (Fig.~\ref{res:4cubes}).
        We built four new cubes in this way, with the same size as the original data cube, by interpolating among the chosen points. 
        
        We ran the regularized 3D fitting method on each of these new data cubes and obtained four maps for each Gaussian parameter. 
        We calculated the resulting standard deviation and divided it by 2 (because the information was now four times less significant). 
        The result is a map of uncertainty for each Gaussian fitting parameter.
        
        In addition to the above described statistical errors, some systematic errors are attached.
        \begin{itemize}
        \item[--] The decision of which hyperparameters to inject into the algorithm as well as the initial guess for the fit procedure influences the resulting maps, especially where the S/N is poor.                
        When the method works properly, parameter maps are smooth and show significant spatial variations. 
        This means that the model resembles the observations. In this case, small variations of hyperparameters lead to variations in the parameter maps, which remain within the previously discussed error bars. 
        Moreover, changes occur more frequently in the values (with a global scaling factor) than in the shapes, which are stable. 
        When the criterion reaches an absolute minimum, the result becomes stable.         
        This source of error is therefore not significant compared to the statistical uncertainties.
        \item[--] Another source of error is the continuum estimation. Continuum is not the same everywhere and is evaluated on some portions of the spectrum that are free of any gaseous or stellar line. 
        However, some H$_{2}$ lines lie very close to stellar lines. 
        The continuum is estimated by a polynomial fit and subtracted before the regularized 3D fitting algorithm is applied. 
        Because of poor S/N, it is sometimes difficult to determine where the continuum is standing. 
        This uncertainty does not affect the parameter maps we built for each line too much. 
        However, the relative strength of the different lines can be affected, and this problem is taken into account in Sect.~\ref{ml1D}.
        \item[--]As previously discussed, the absolute flux map has one additional source of error that comes from the photometric calibration (Sect.~\ref{cal}). 
        \end{itemize}

                \begin{figure*}
                \centering
                \vspace{-0.5cm}
                \includegraphics[width=9.1cm]{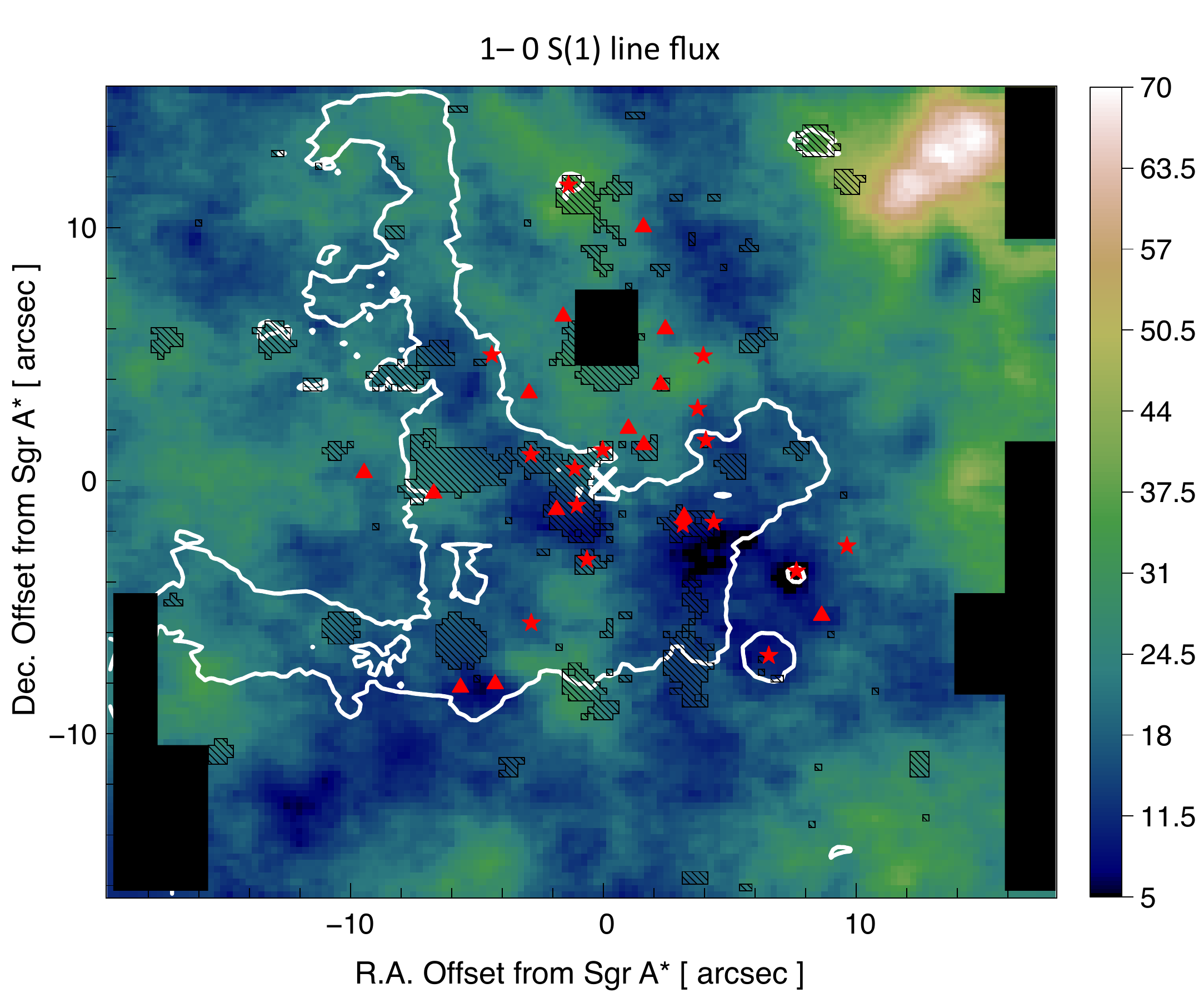}
                \includegraphics[width=9.1cm]{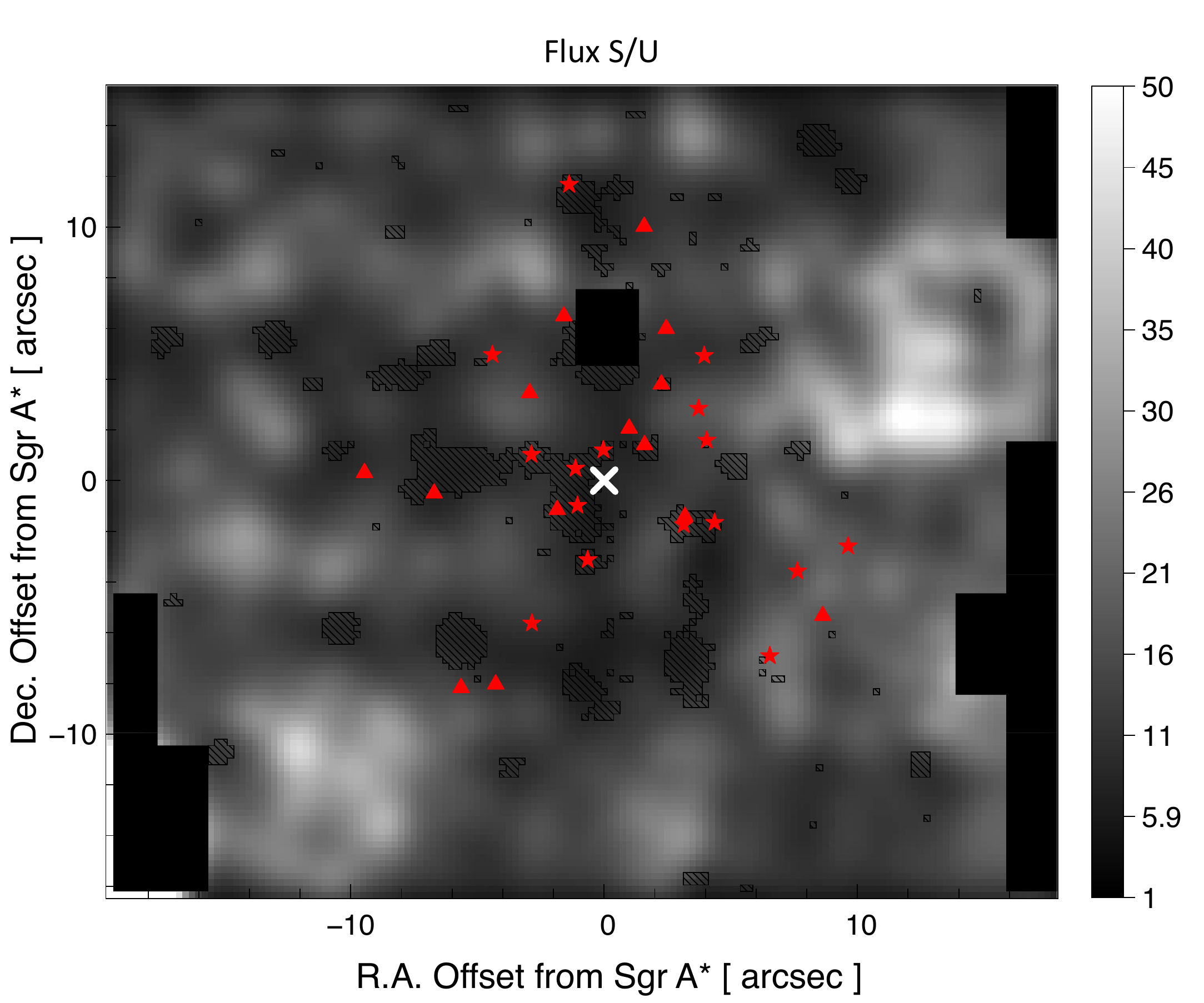}\\
                \includegraphics[width=9.1cm]{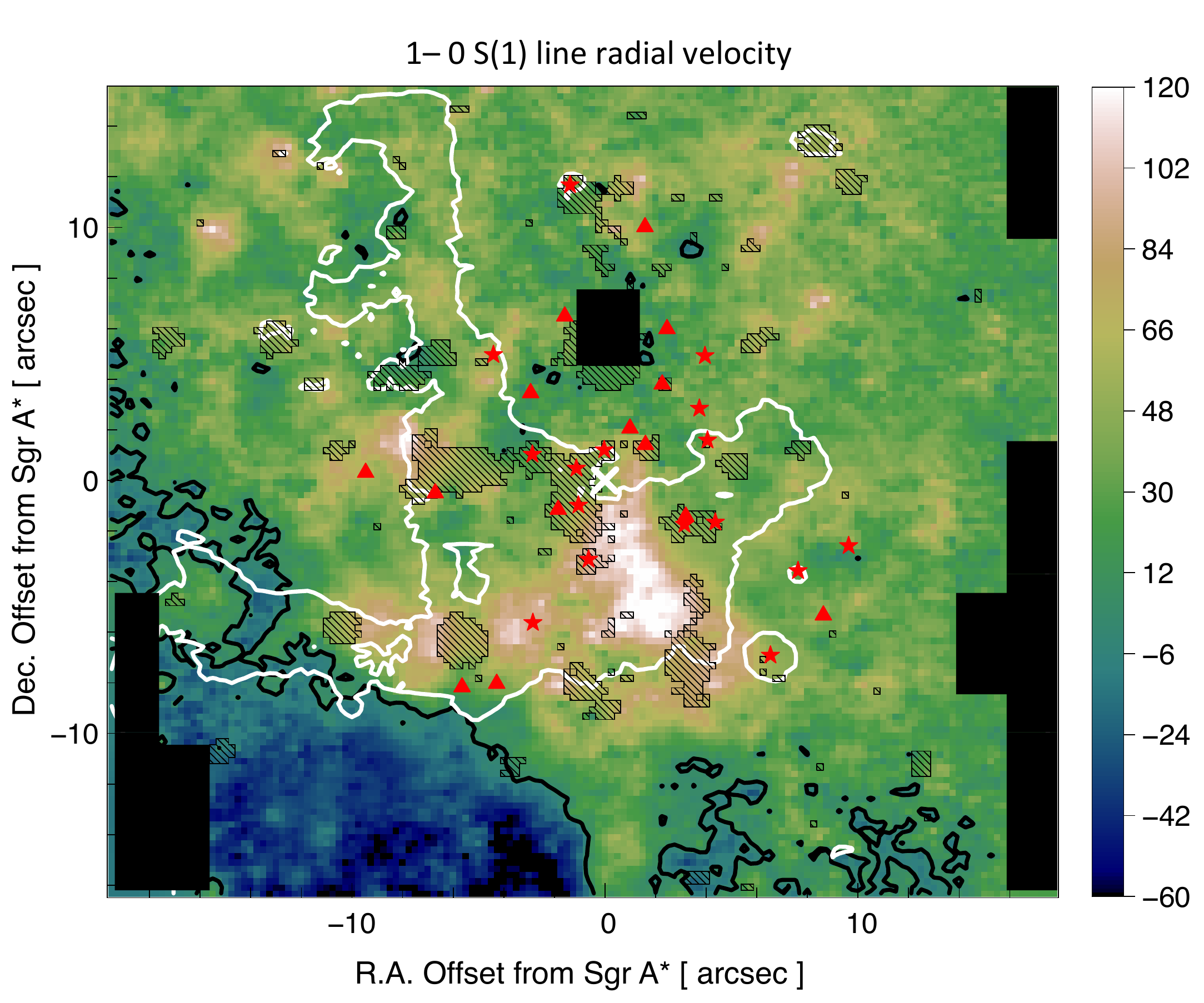}         
                \includegraphics[width=9.1cm]{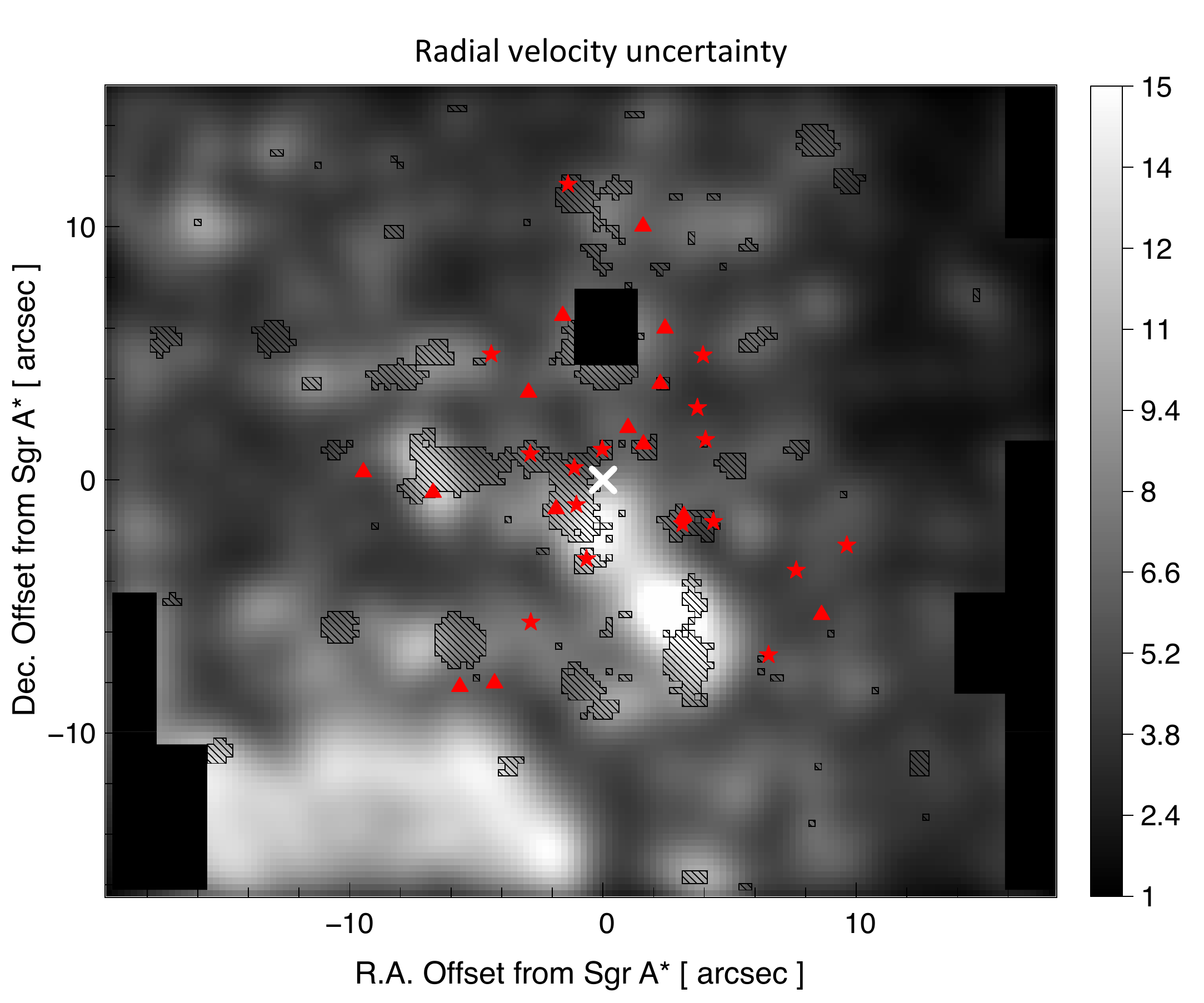}\\
                \includegraphics[width=9.1cm]{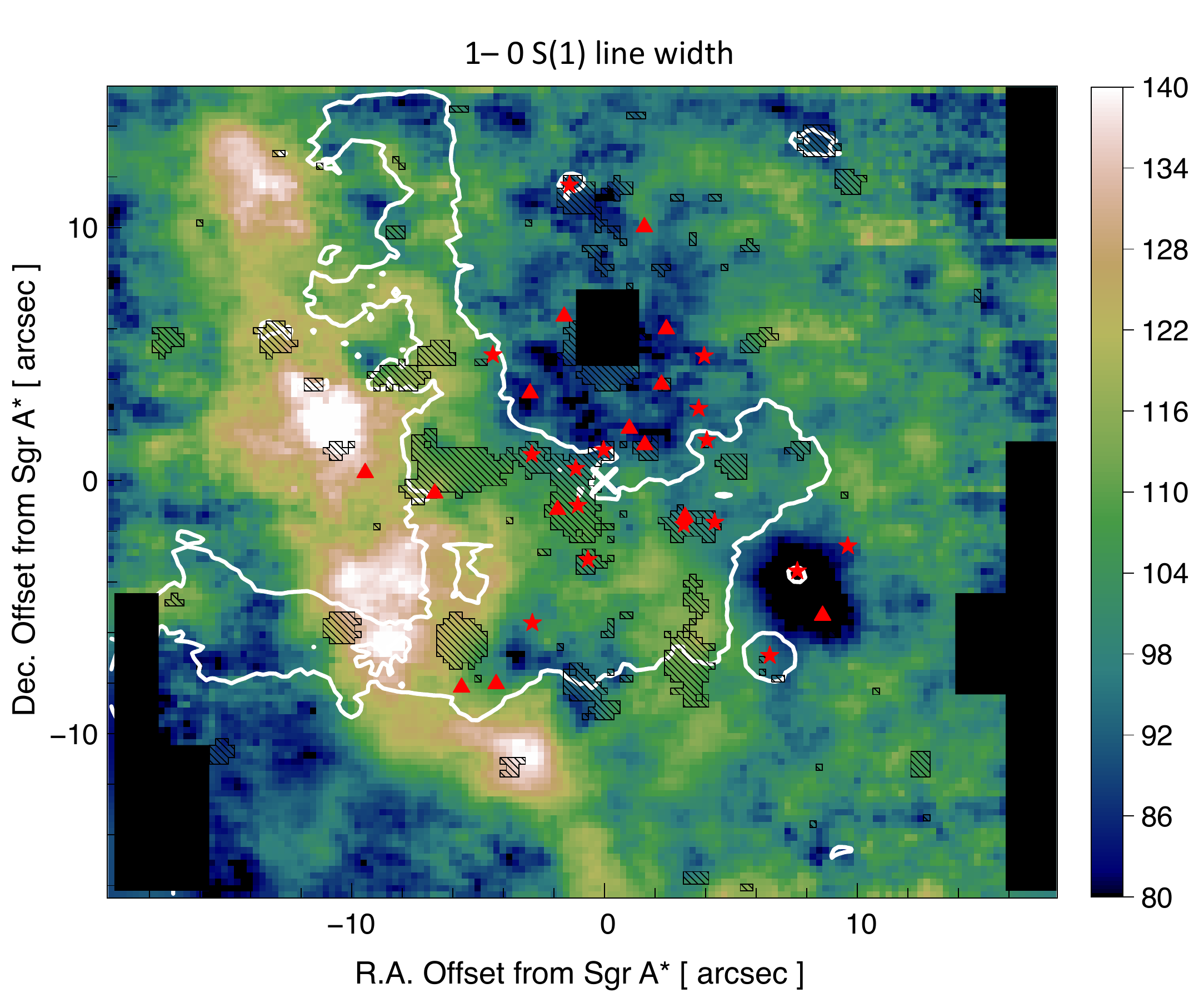}
                \includegraphics[width=9.1cm]{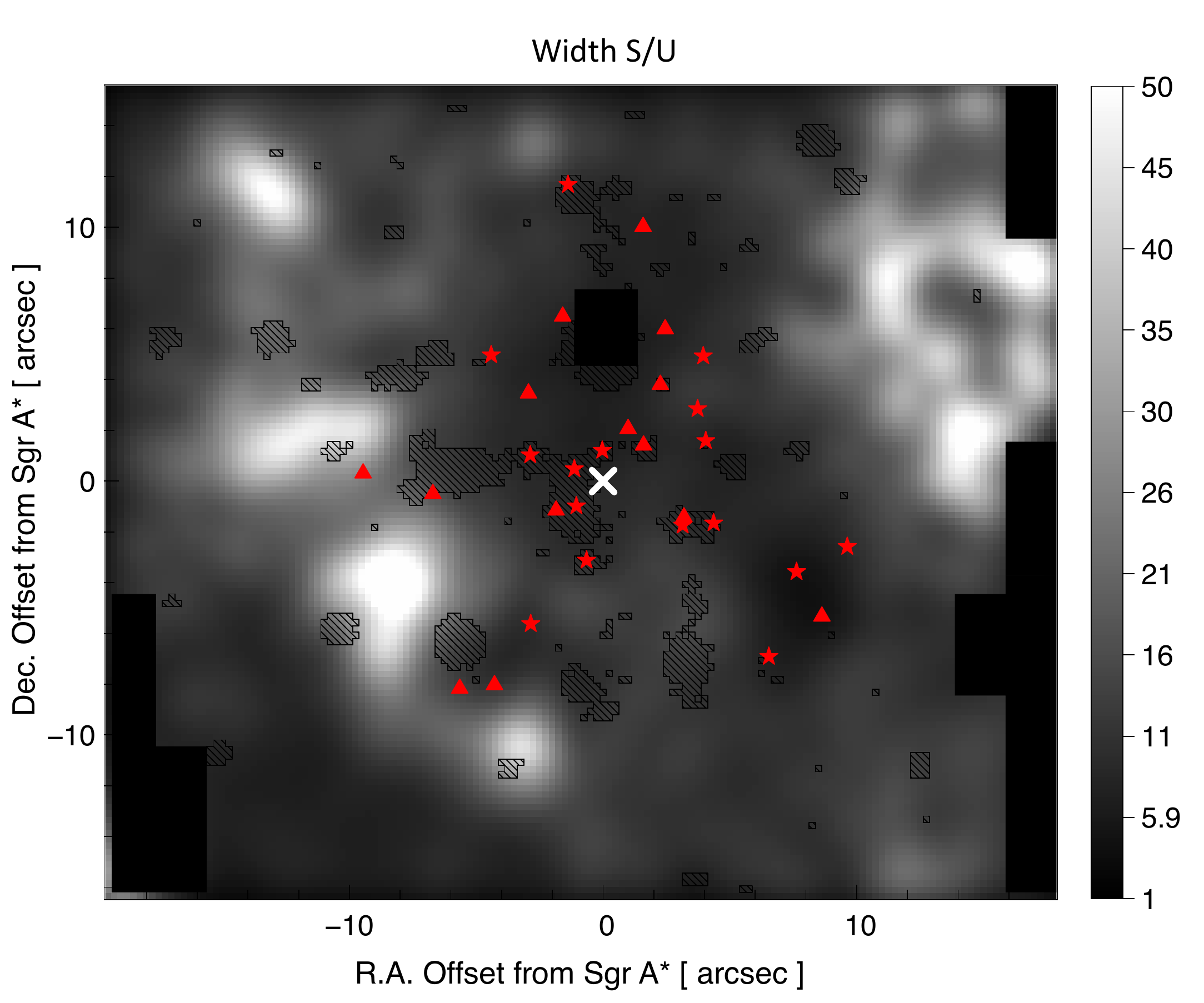}
                \caption{\textit{Left:} Parameter maps of the 1--0~S(1) line obtained through regularized 3D fitting. 
                Sgr~A* is indicated by the cross. The line flux color bar is in 10$^{-19}$~W~m$^{-2}$~arcsec$^{-2}$. 
                Velocity and width color bars are in k~s$^{-1}$. 
                The white contour traces the Br$\gamma$ emission of the minispiral at 2$\cdot$10$^{-16}$~W~m$^{-2}$~arcsec$^{-2}$. 
                The dashed areas cover stars zones that have been interpolated. 
                Red dots indicate the position of Wolf-Rayet stars: stars show  type WN and triangles type WC (\citealt{Paumard06} catalog). 
                The solid black areas indicate regions of which we lack observations. 
                GCIRS~7 is covered by the central black square. The solid black line in the velocity maps indicates points at zero velocity. 
                \textit{Right:} signal-to-uncertainty ratio (S/U) maps for flux and width maps, uncertainty map for velocity.}
                \label{res:maps}
                \end{figure*}

                \begin{figure*}[htbp]
                \begin{center}
                \includegraphics[width=17.9cm]{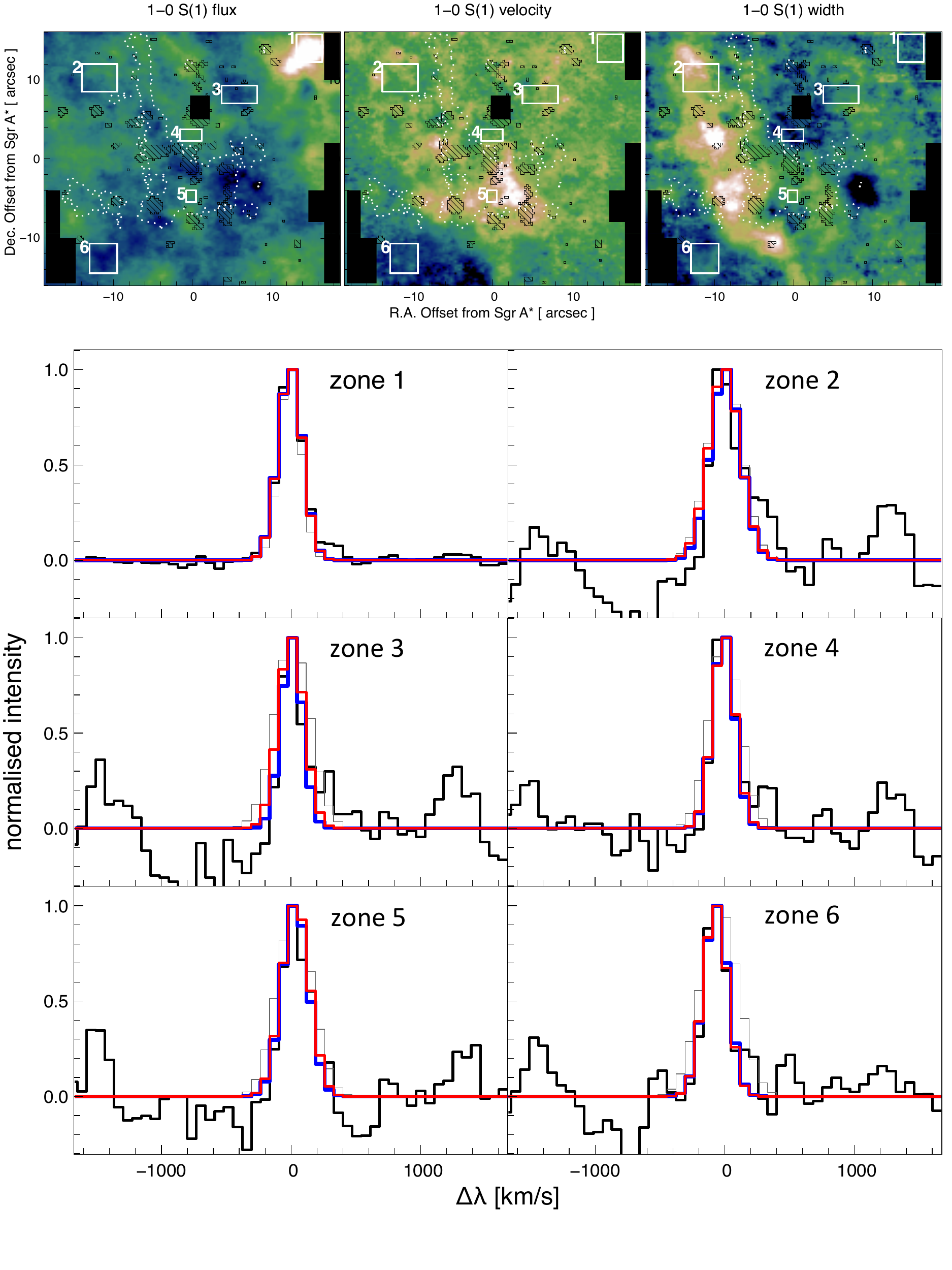}
                \vspace{-1.2cm}
                \caption{\textit{Top:} Zones where individual spectra of S(1) line, have been extracted, superimposed on parameters maps (the same as in Fig.~\ref{res:maps}). 
                \textit{Bottom:} for each zone the normalized average spectrum (black) is superimposed on the respective pixel by pixel 1D model (solid blue) and regularized 3D model (solid red). 
                The gray line reproduces another solution of the regularized 3D fitting. 
                Both solutions have close values of $\chi2$ (1.15 and 1.49, respectively), but the red model fits the observations
much better.}
                \label{res:zones}
                \end{center}
                \end{figure*}
                                
\subsection{Regularized 3D fit maps}
\label{reg}

We started our analysis with the H$_2$ 1--0~S(1) transition line at 2.1218~$\mu$m, which has the best S/N. 
As previously mentioned, a pixel-by-pixel fit (1D fit) was performed on some zones of the field to be compared with the regularized 3D fit and verify the method.

        \subsubsection{1--0~S(1) line }
        \label{3d1d}
        
        The retrieved maps of line flux, radial velocity, and width are reported in Fig.~\ref{res:maps} with the respective error maps. \\
        
        \noindent \textit{Flux map}\\ 
        
        The flux map shows that the molecular gas is detected in the whole field of view.       
        Some of the most prominent features are 
        \begin{itemize}
        \item[--] a bright emission at the CND position in the northwestern corner;
        \item[--] a weak emission situated between two stronger emissions in the southeastern corner, where the CND lies as well;
        \item[--] a plume-like feature in the north near GCIRS~7
to the right of the minispiral northern arm;
        \item[--] in the southeastern corner we recognize a feature known as the southern extension \citep{Christopher05}.
        \end{itemize}
                
         \newpage
         \noindent \textit{Velocity map}\\

        The velocity map presents a well-pronounced blueshift in the southeastern corner, which is consistent with the CND motion toward the observer. 
        The remaining velocity map has a uniform global pattern (moving at around 30~--~50~km~s$^{-1}$) and several small-scale structures. 
        They mostly correspond to bright stars. 
        These features can be artifacts created by the stellar pollution, but they may also be associated with gas ejected by the stars themselves. 
        The velocity at the center of the map is strongly affected by nearby stars too. 
        Here the S/N is low (1 or 2). 
        However, this location is very close to the minicavity, a shell of $\sim5\arcsec$ that is probably created by the interaction between the stellar and the interstellar medium \citep{Eckart92}. 
        It is a region of shock. 
        The high velocity we observe here may be related to this shock. 
        A more detailed analysis of this area will be covered in future works with a second dataset that has a better spatial resolution. \\
        
         \noindent \textit{Width map}\\
        
        The H$_2$ 1--0~S(1) width map presents some enhancements (i.e., large width) along a ridge that is slightly shifted to the east of the northern arm. 
        This arm is a thick cloud (hereafter the northern arm cloud; \citealt{Jackson93, Paumard04}) whose eastern boundary is ionized and traced by the north-south Br$\gamma$ emission. 
        The width map features correspond in projection to the denser part of this cloud, suggesting that the H$_2$ gas inside is more turbulent. 
        The H$_{2}$ 1--0~S(1) line looks narrower in the southeastern and northwestern corners, at positions associated with the CND. 
        The line is wider in the central cavity, with a few exceptions (the area around GCIRS 7 and a region 8\arcsec~west and 4\arcsec~south of Sgr A*). 
        We would like to point out that the line width here represents the fitted measured width, which is the quadratic sum of intrinsic and instrumental width. \\ 
        
        \noindent
        To assess these results, we extracted individual spectra in sixteen zones and fit them with a classical 1D model. 
        In the remaining paper we concentrate on six of the zones that are most representative of peculiar areas (CND, minispiral, central cavity, and minicavity). 
        The detailed information on the other zones is given in Appendix~{\ref{A}}.  
        We compared this 1D model to the result of the regularized 3D fit (Fig.{\ref{res:zones}}). 
        The two models can hardly be distinguished, which shows that when it is integrated over an aperture, our 3D method is equivalent to classical data analysis.
                                                
        \subsubsection{Other 1--0~ortho lines: S(3), Q(1), and Q(3)}
        \label{orthomaps}
        
        We have applied the fitting method to the other ortho lines available in our dataset. 
        The transitions with a sufficient S/N are 1--0~S(3), 1--0~Q(1), and 1--0~Q(3) at 1.9575~$\mu$m, 2.4066~$\mu$m, and 2.4237~$\mu$m.
        
        These lines lie in a more complex spectral environment than the S(1) line. S(3) is near the atmospherically absorbed range between the H and K band. 
        Q(1) and Q(3) are close to CO stellar photospheric absorption features (see Fig.~\ref{res:orthoSP}). 
        The retrieved maps for each of these lines are displayed in Fig.~\ref{res:orthoMaps} (second, third, and fourth columns). 
        
        Even though the fitting procedure is more challenging than for 1--0~S(1) maps, the three ortho line maps are fairly similar to those of 1--0~S(1). 
        The flux maps in particular show the same patterns. 
        The velocity maps also have the same gradient shape. For the width maps we note the following.
        \begin{itemize}
        \item[--] For 1--0~Q(3) the  width map confirms the previous finding of an emission from the northern arm cloud, at least in projection.
        In the southeastern corner of the map the line appears wider than 1--0~S(1). However, the width map structure globally
presents the same peculiarities as 1--0~S(1). 
        \item[--] 1--0~S(3) is affected by a close atmospheric line. This probably causes the higher values of the width, but the overall shape of the width map is compatible with the one obtained with S(1). 
        \item[--] The Q(1) line is affected by the nearby CO features. This absorption probably reduces the measured flux and explains the lower values of the width. 
        \end{itemize}

        In summary, when the local peculiarities and spectral pollution are taken into account, the parameter maps of these four ortho H$_{2}$ lines are compatible. 
        This strengthens the validity of our results.
        Indeed, if all the lines were emitted by a common source, it is expected that the same dynamics applies for all H$_{2}$ molecules, regardless of their excitation state. 

        According to this interpretation, all lines can be fit using the velocity and width obtained for 1--0~S(1). These parameters are best estimated for this line. 
        Hence we applied the regularized 3D fit procedure a second time by imposing these conditions on the other lines (1--0~S(3), Q(1) and Q(3)). 
        We obtained new flux maps (Fig.~\ref{res:orthoMaps}, first column).
        These flux maps with an imposed velocity and width are very similar to the fit where all parameters were let free to
vary.
        This confirms the relevance of using a common width and velocity.
        This assumption improves the fitting, since two of the parameters rely on the most reliable line, which decreases the effect of spectral environment pollution.

        \subsubsection{Para 1--0~S(0) line}
        \label{para}

        The 1--0~S(0) line is the para line with the best S/N. 
        However, there is a \ion{Fe}{iii} line on one side of the H$_{2}$ line and an absorption feature on the other.
        Despite these difficulties, S(0) is clearly detected almost everywhere in the SPIFFI field of view. 
        However, when we tried to fit S(0), the regularized 3D
fit did not converge on a satisfactory solution. 
        A satisfactory fit can be achieved only if the line is simultaneously fit with 1--0~S(1). 
        In this case, the model fits both lines through four parameters: S(1) flux, S(0) flux normalized by S(1) flux, velocity, and width.

                \begin{figure*}[htb]
                \centering
                \includegraphics[width=17cm]{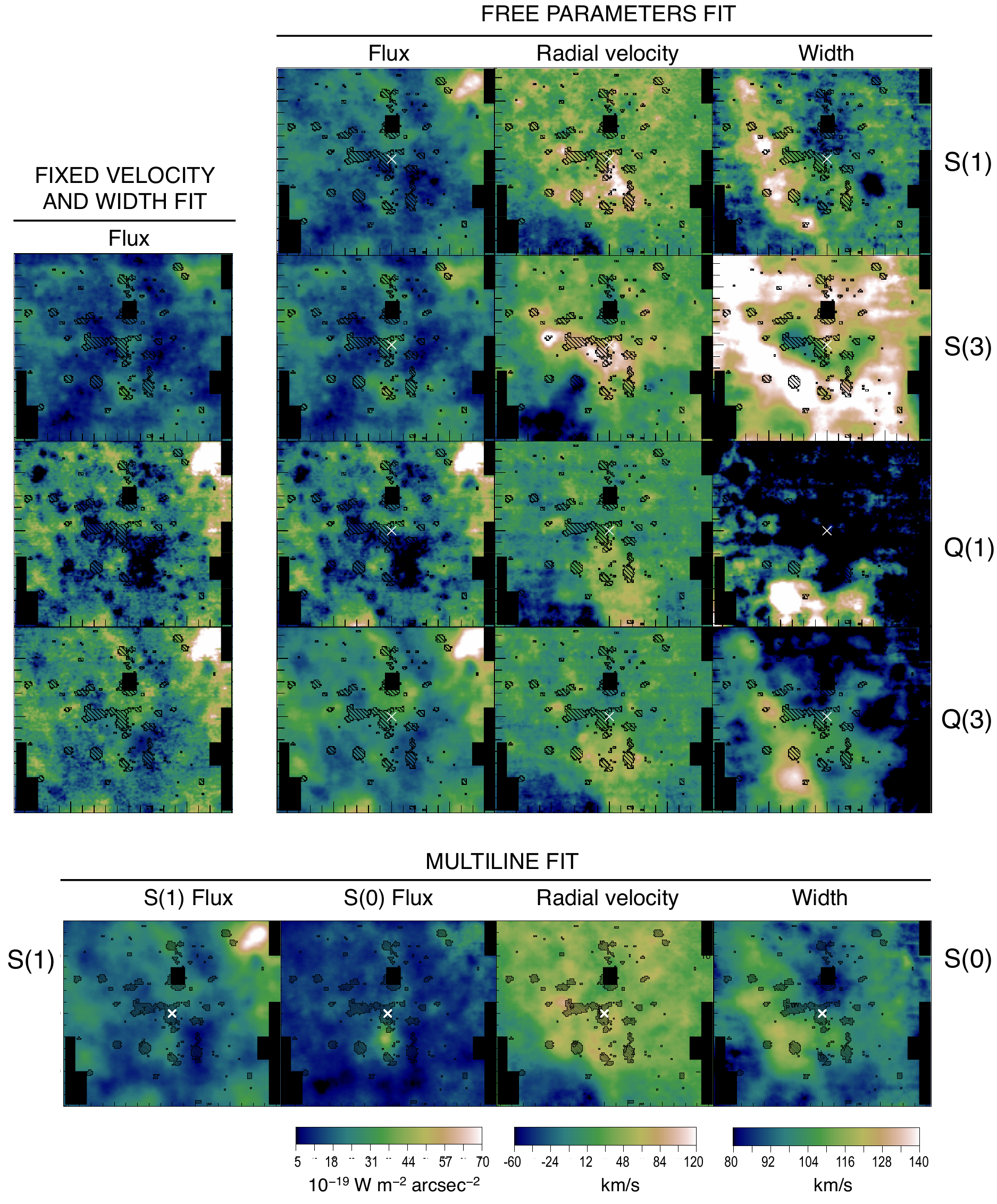}         
                \caption{Parameter maps for each ortho line. \textit{Left to right:} flux with velocity and width imposed to be those of S(1) line. 
                Line flux, velocity, and width with free parameters.
                \textit{Top to bottom:} 1--0~S(1), S(3), Q(1), and Q(3) lines plus the 1--0~S(0) line maps obtained with the multiline fit. 
                Maps are centered on Srg A*. 
                The color bars are the same for all lines and as in Fig.~\ref{res:maps}. 
                Symbols as in Fig.~\ref{res:maps}.}
                \label{res:orthoMaps}
                \end{figure*}

        Figure~\ref{res:orthoMaps} displays the parameter maps obtained for this simultaneous fit. 
        The S(1) flux, velocity, and width maps are compatible with the previous findings. 
        The S(0) para line is therefore well described by the same velocity and width as S(1) and other ortho lines. 
        The main difference with respect to other studied lines is that this transition shows a strong emission maximum a few arcseconds south of Srg A*. 
        In this zone, the S/N is good and the S(0) line very clearly stands out from the continuum. 
        This peak of 1--0~S(0) emission is discussed in detail in Sect.~\ref{prop}.
        
\subsection{Extinction-corrected map}
\label{Schodel10}

Extinction correction is crucial not only to obtain the absolute flux values, but also to compare fluxes in different lines (Sect.~\ref{prop}).
We considered the 1--0~Q(3) to 1--0~S(1) flux ratio for the extinction evaluation because these two lines correspond to the same upper level. 
We applied regularized 3D fitting procedure directly on this ratio, instead of fitting the two lines separately and considering the ratio afterward, as for the S(0) to S(1) ratio (Sect~\ref{para}). 
Fitting the flux ratio directly allows the result to be less affected by small-scale artifacts and local variations of any of the parameters. 
        
Since the H$_{2}$ lines are optically thin (as shown in the Orion nebula, for instance, by \citealt{Gautier76}) we can write
\begin{equation}
\frac{I_{Q3}}{I_{S1}}=\frac{(\mathcal{A}_{Q3}\cdot \nu_{Q3})}{(\mathcal{A}_{S1}\cdot \nu_{S1})}\cdot 10^{E/2.5}
,\end{equation}
where I$_{Q3}$/I$_{S1}$ is the intensities ratio (i.e., the flux ratio), $\mathcal{A}$ is the Einstein coefficient and $\nu$ the frequency. 
Considering $E=A_{\lambda_{S1}}-A_{\lambda_{Q3}}$, we have
\begin{equation}
E=2.5 \cdot log \left( 1.425\cdot \frac{I_{Q3}}{I_{S1}}\right)
\label{E}
.\end{equation}
The 1.425 factor in Eq.~\ref{E} comes from the ratio of the Einstein coefficient of the two transitions. 

Assuming an extinction law, we can obtain the total extinction A$_{\lambda}$ at every wavelength  and correct every line at every pixel for reddening. 
We assumed a power-law in the NIR as in Sect.~\ref{relcal}. 
In particular,  we obtain a map for the extinction at Ks through the equation $A_{a}/A_{b}=(\lambda_{a}/\lambda_{b})^{-\alpha}$ (with $\alpha=2.07$ from \citealt{Fritz11}), where $\lambda_{Ks}=2.168$~$\mu$m as in \cite{Schodel10}. 

\begin{figure}[htb]
                \centering
                \includegraphics[width=9cm]{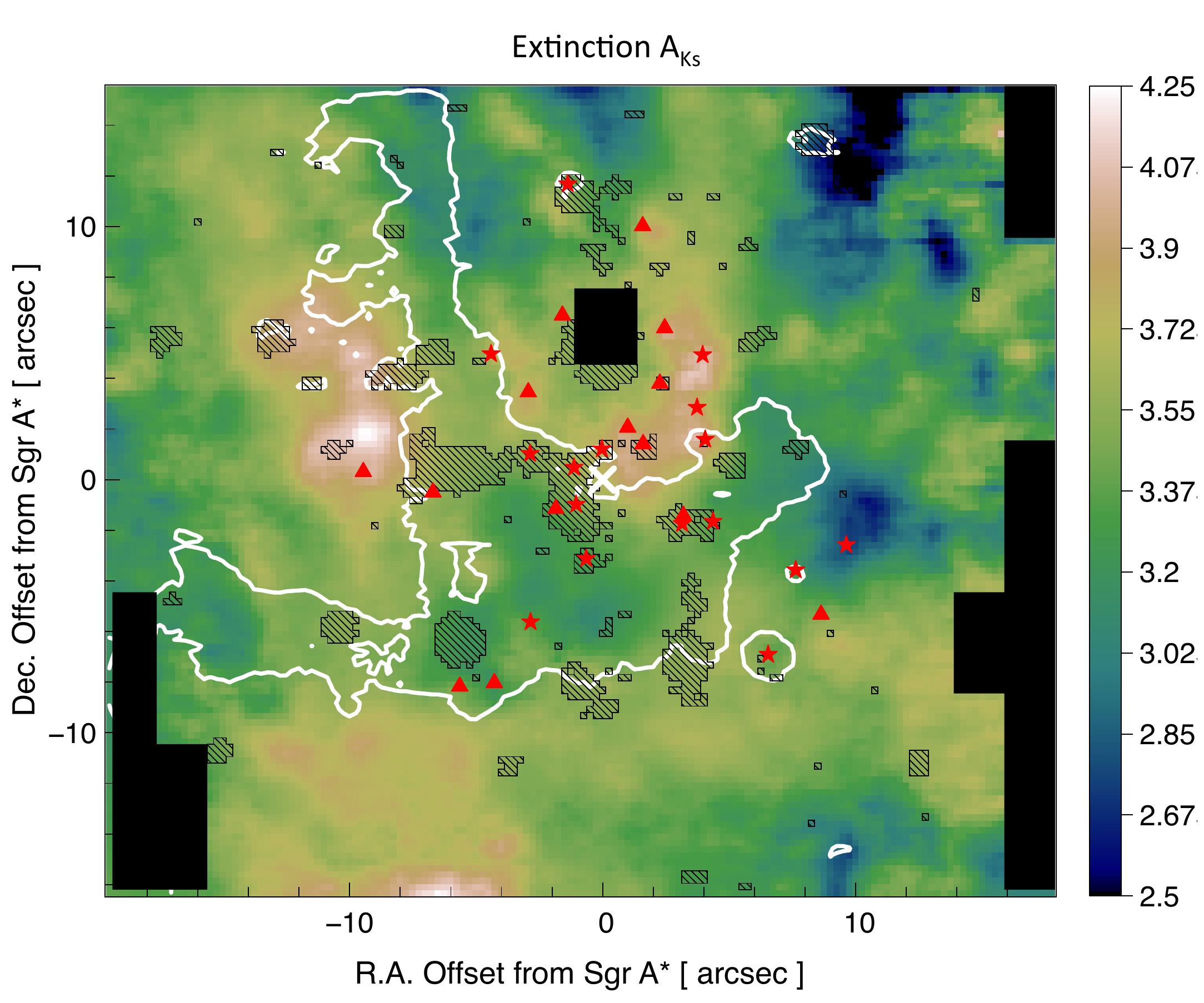}\\
                \includegraphics[width=9cm]{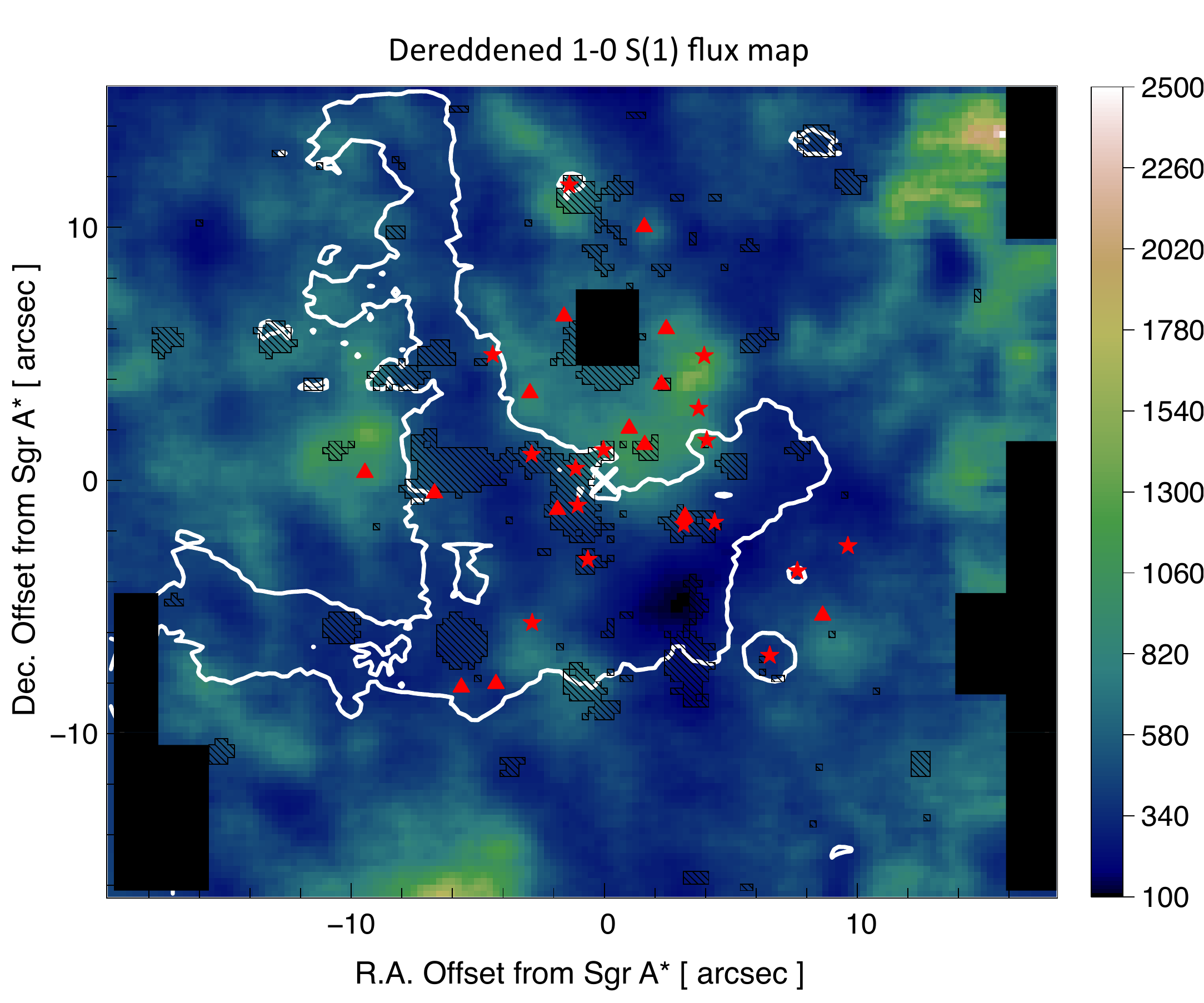}
                \caption{\textit{Top:} extinction A$_{Ks}$ map at $\lambda_{Ks}\sim2.168$~$\mu$m in mag obtained from Q(3) to S(1) ratio. 
                \textit{Bottom:} S(1) flux map, dereddened through this extinction map. 
                Symbols as in Fig.~\ref{res:maps}.}
                \label{s1ml}
                \end{figure}

\begin{figure}[htb]
                \centering
                \includegraphics[width=9cm]{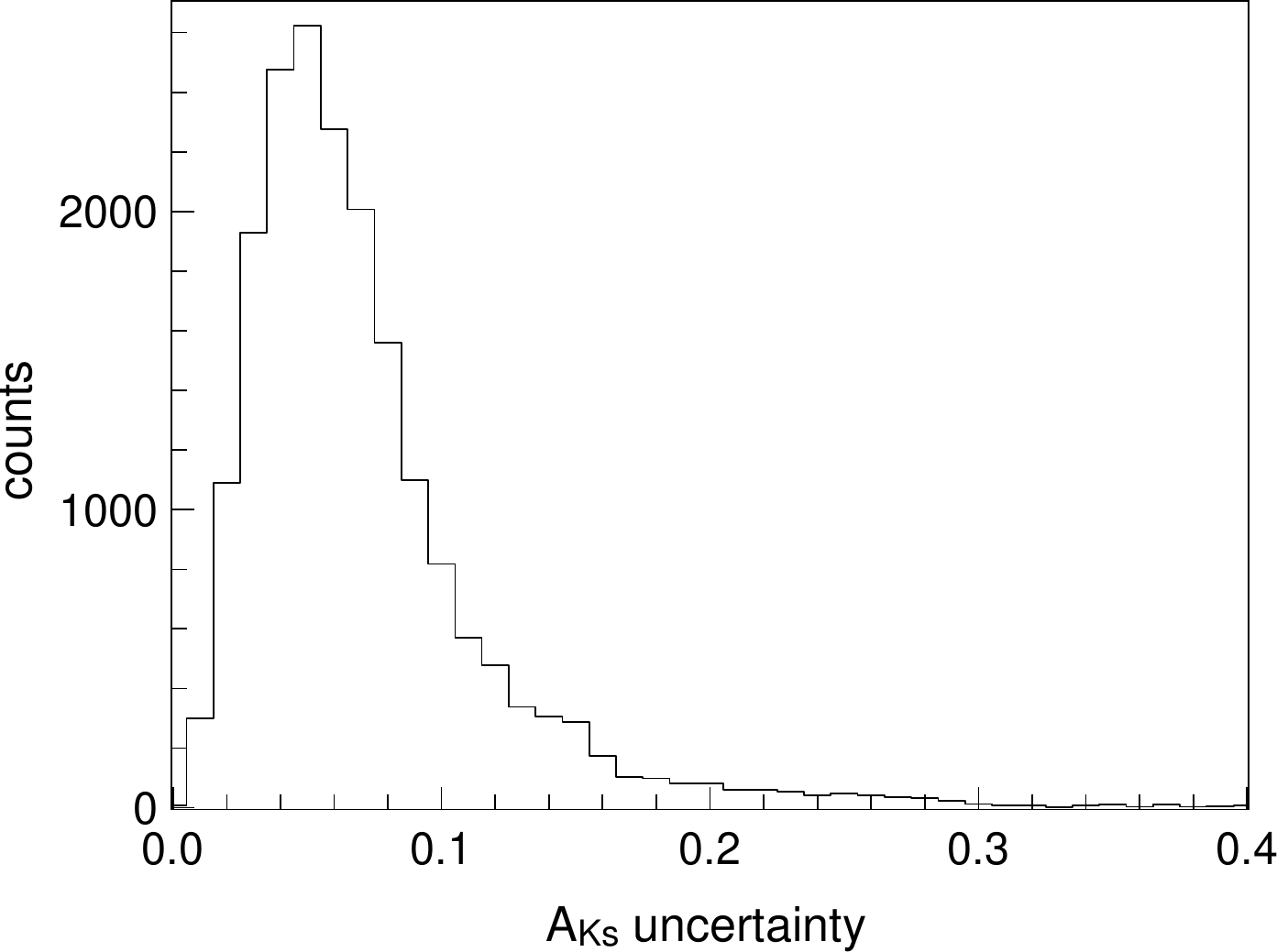}
                \caption{Histogram showing the distribution of the statistical uncertainty of A$_{Ks}$.}
                \label{histo}
                \end{figure}

The A$_{Ks}$ map, together with the corrected flux map of S(1), is reported in Fig.~\ref{s1ml}. 
We computed the uncertainty on the extinction from the propagation of Q(3) to S(1) ratio uncertainty, obtained as explained in Sect.~\ref{EB}. 
The histogram of the resulting statistical error is reported in Fig.~\ref{histo}. To this value a systematic error of 10\% needs to be added because of the calibration.

The average value of A$_{Ks}$ is $\simeq$~3.4. The extinction map shows significant variations, well above 1 mag.

The dereddened 1--0~S(1) flux map shows essentially the same global features as the uncorrected map. 
However, the western arc and the southern extension emissions are enhanced. 
The emission is
also enhanced between GCIRS~7 and the bar of the minispiral. 
This feature is also present, albeit less strikingly, in the uncorrected map (Fig.~\ref{res:orthoMaps}).

The dereddened flux map appears partially correlated with the extinction map: the flux is high where the extinction is high as well. 
A more detailed analysis of this correlation is the object of another paper (\citealt{Ciurlo15} and Ciurlo \textit{et al.} 2016, in prep), where we model this effect.  

\subsection{Discussion}

        \subsubsection{CND emission}

        The emission observed in the northwestern and southeastern corners of the SPIFFI field of view is most probably dominated by CND molecular gas that is located there (flux map of Fig.~\ref{res:maps}). 
        Because the CND is inclined with respect to the plane of the sky \citep{Liszt03}, the CND western border (in the northwest of the SPIFFI field) is closer to the observer. 
        This is coherent with a stronger emission from the closest, and thus less extinct, side (the northwestern corner).

        The flux map we obtain shows the same features as the 1--0~S(1) map obtained by \cite{Yusef-Zadeh01}.  
        For instance, both maps show the bright emission in the northwestern corner and the weak emission in the opposite corner (with the same structure of a gap between two brighter features) as well as the plume (which is probably not associated with the CND, however). 
        The flux associated with these regions is also compatible with the values obtained by \cite{Yusef-Zadeh01}. 
        For instance, in the northwestern corner we found an undereddened flux of $\sim$~60~$\pm$~10~10$^{-16}$~erg~s$^{-1}$~cm$^{-2}$~arcsec$^{-2}$,  while \cite{Yusef-Zadeh01} found $\sim$~50~10$^{-16}$~erg~s$^{-1}$~cm$^{-2}$~arcsec$^{-2}$ with an uncertainty of 30\%.  
        We consider this agreement as another confirmation of the validity of our method. 

        Some features are also compatible with the HCN distribution obtained by \cite{Christopher05}. 
        HCN clearly traces the densest clumps in the CND. 
        The southern extension is particularly evident. 
        The northwestern corner and western border also coincide with the HCN distribution. 
        This is another strong indication that all these emissions come from the CND itself.

        \subsubsection{Central cavity emission}

        It is interesting to note that the H$_2$ emission remains fairly strong inside the CND borders. 
        This emission could arise either from the background or from the central cavity itself, close to the center. 
        In the central cavity H$_{2}$ is thought to resist dissociation with
a far lower probability.
        However, in the following sections we provide indications that this emission is indeed in the central parsec.
        
        We note that the 1--0~S(0) line exhibits 
        a strong maximum near Srg~A*, at the entrance of the minicavity. 
        This feature could result from a shock, which is not unusual in this region, as previously discussed, or from some particular environmental conditions. 
        This point is discussed in more detail in Sect.~\ref{prop}.
        
        Another flux enhancement is visible between the minispiral bar and GCIRS~7.
        This flux feature is very well correlated with the position of some Wolf-Rayet stars of type WC (shown by the triangles in Fig.~\ref{s1ml}). 
        The environment of these stars is often dusty. 
        Some dust has been detected by \cite{Tanner02} and \cite{Haubois12} at these locations, which means that forming H$_{2}$ is expected. 
        These stars may be located in the background, farther from the central cluster than the northern arm. 
        This would explain why the outer layer of these dust shells is not ionized and consequently not observed in the ionized gas maps.
        
        \subsubsection{Gas dynamics}
	\label{gasdyn}
	
        The velocity map we obtained shows that the northwestern corner and western border are moving at $\sim$~30~--~50~km~s$^{-1}$. 
        These values are well compatible with the HCN observations: \cite{Christopher05} found $\sim$~50~km~s$^{-1}$ and 50~--~60~km~s$^{-1}$ , respectively. 
        The velocity corresponding to the above-mentioned southern extension is around $-$40~km~s{$^{-1}$}. 

        We observe no large-scale motion of H$_{2}$ associated with the minispiral arms. 
        Even though at least part of the emission seems to be connected to the minispiral (the northern arm cloud), we see no evidence of a transport of matter along its arms.

        The western part of the field that corresponds to the CND border looks smoother in velocity than the central cavity. 
        The reason is that the central cavity has more inhomogeneous dynamics
than the CND. 
        
        The spatial distribution of the width is very different from the flux map. 
        Almost throughout the central cavity, the width map shows a wider but weaker line emission, with a particularly strong width enhancement corresponding to the northern arm cloud. \\
         
        \noindent        
        In summary, there seem to be at least two different emission
components:
        \begin{itemize}
        \item[--] one associated with the CND (northwestern and southeastern corners mainly) that presents a narrow and intense emission line, and
        \item[--] one inside the central cavity (at least in projection) that is generally characterized by a wider but weaker line emission than the other component.
        \end{itemize}
        In the CND the line is narrower than in the remaining field probably because its clumps are
more dense, which prevents the line from integrating deep into them. 
        On the other hand, the gas is less dense in the central cavity. 
        Here, the observed widening of the line translates the dispersion in velocity along the line of sight. 
        This effect is particularly strong inside the thick northern arm cloud.
        
        \subsubsection{Extinction map}

        The average extinction we derive is almost 1 mag greater than that determined by \cite{Schodel10} through star color excess in the same region. 
        \cite{Fritz11} computed the extinction curve toward the Galactic Center using the same SPIFFI data cube we analyzed and found results in agreement with \cite{Schodel10}. 
        However, they used ionized hydrogen lines to calculate the extinction.
        These differences are therefore expected. 
        
        The bulk of extinction is in the foreground and amounts to $\sim$~2.5 K mag, but most of the variation in extinction is local: up to 1 additional magnitude at K. 
        For instance, \cite{Paumard04} showed that the dust contained in the eastern bridge is responsible for the 0.76 mag extinction of the northern arm at K. The 
        \cite{Schodel10} A$_{Ks}$ map shows local variations of more than 1 mag as well. 
        However, our map shows even more variations. 

        The maps also shows different features.
        For instance, \cite{Scoville03} observed the Pa$\alpha$ and 6 cm radio continuum emission of the central few parsecs.
        The extinction map built by \cite{Scoville03} exhibits a minimum along the northern arm cloud. 
        In contrast, the extinction map we obtain shows a maximum corresponding to the same cloud. 
         An anticorrelation seems to lead to the interpretation that we are indeed observing an emission from inside the cloud, whereas the extinction inferred by \cite{Scoville03} concerns the surface of the same cloud.

        Our interpretation is that the ionized gas is essentially sensitive to the foreground extinction because it is situated at the surface of clouds, while molecular gas is located more deeply inside a clump. 
        Therefore H$_{2}$ emission is more strongly affected by the local variable extinction.
        
        Extinction estimation depends on the 3D position of the object that is used as a probe. 
        Extinction maps may differ because the various tracers,
stars, ionized gas, and molecular gas, are at different 3D positions and different optical depths. 

\section{Excitation of the molecular hydrogen}  
\label{prop}

We now investigate the physical properties of the observed H$_{2}$ emission in the central parsec.
The relative line intensities give information on the gas excitation temperature, while the absolute flux values allow determining the total column density as well as the mass of excited H$_{2}$. 
Therefore, the comparison of several lines can reveal different specific processes and give important indications on the dominating excitation mechanism. 

For the comparison of intensities required by the study of the H$_{2}$ physical conditions, it is important to gather as many lines as possible to obtain stronger constraints on the gas properties.
In addition to the previously discussed lines, some other lines are detectable: the 1--0~S(2), Q(2), and 2--1~S(1) lines at  2.0338~$\mu$m, 2.4134~$\mu$m, and 2.2476~$\mu$m. 
\begin{itemize}
\item[--] The Q(2) line is strongly affected by CO features and it is safely detected only in the northwestern corner, where the CND lies and where all detected lines show a maximum. 
\item[--] The S(2) line is probably polluted by one or more nearby OH sky lines that are not perfectly subtracted. However, it is detectable in the northwestern half of the field.
\item[--] The 2--1 S(1) line is weaker and therefore the S/N is low. This transition is detected mainly at the border of the field, where the CND lies.
\end{itemize}

\begin{figure}
                \centering
                \includegraphics[width=9cm]{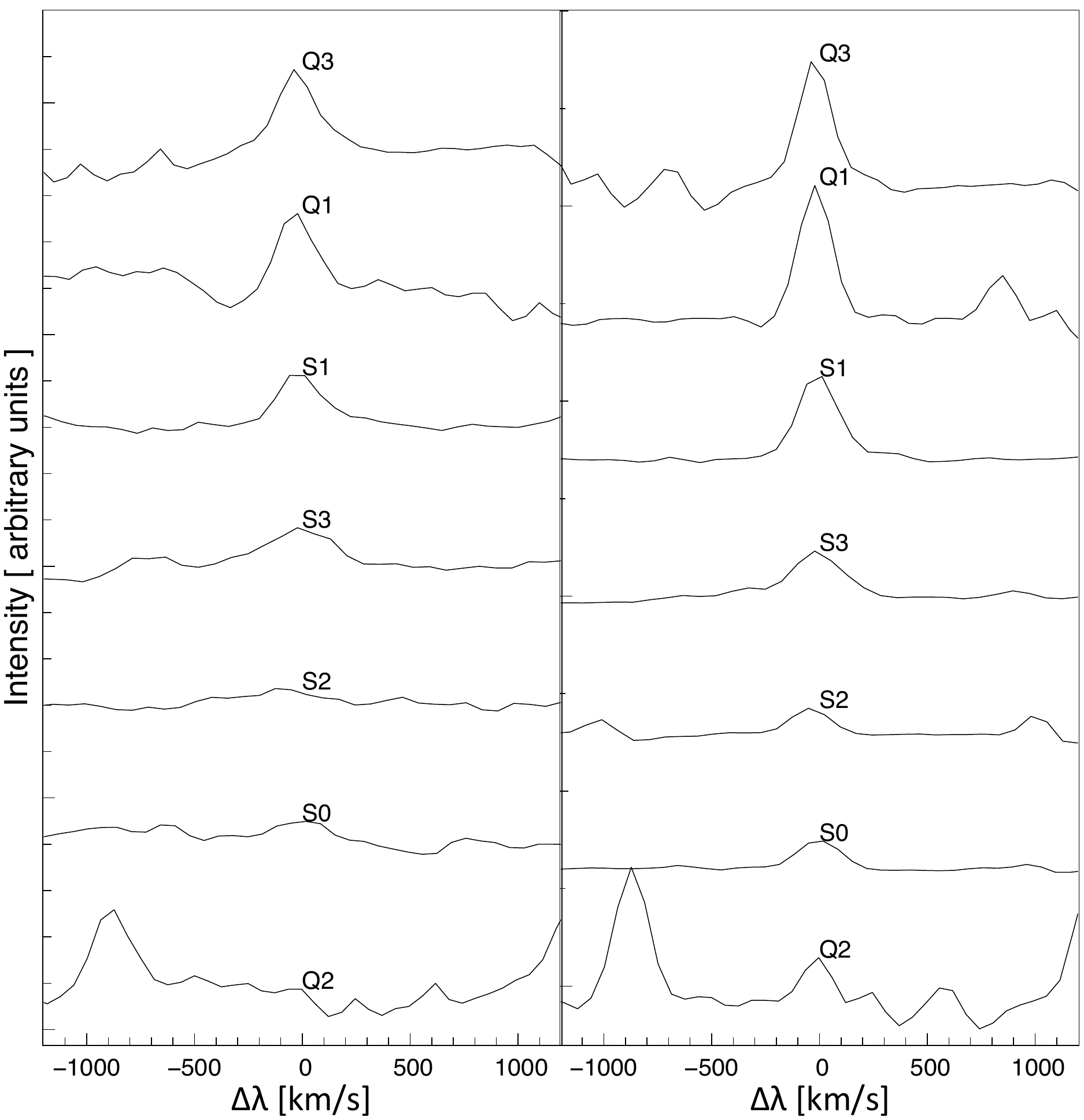}
                \caption{Ortho (\textit{top}) and para (\textit{bottom}) uncalibrated spectra, arbitrarily shifted along the y-axis for the sake of comparison, averaged across the whole field (\textit{left}) and  only over the CND zone 1 (\textit{right}). }
                \label{res:orthoSP}
                \end{figure}

These lines are difficult to analyze with the regularized 3D fitting because of the low S/N and nearby atmospheric and stellar features on the spectra (see Fig.~{\ref{res:orthoSP}}). 
No complete maps can be achieved for all detected lines. 
However, H$_{2}$ conditions can be studied through a classical 1D analysis on some regions of the field, where lines are distinctly detectable.
        
\subsection{Method}
\label{ml1D}

The best choice to correctly compare intensities is to simultaneously fit all the H$_{2}$ lines under study. 
This allows all lines to have the same velocity and width and avoids adding further uncertainties.

        \subsubsection{Multiline 1D analysis}

        We considered a spectral window of about $10\,000$~km~s$^{-1}$ around each detectable line and subtracted the continuum. 
        We extracted average spectra over the zones displayed in Fig.~\ref{res:zones}. 
        The zones were selected with the objective of having as much H$_{2}$ lines as possible. 
        However, not all the same H$_{2}$ lines are detectable in every region. 
        Therefore not all zones present the same number of H$_{2}$ lines finally included in the analysis.

        For each zone, the set of selected lines was fit with a multiple Gaussian profile function. 
        The parameters of this function are the principal line flux (here 1--0~S(1) because it has the best S/N), the flux ratio of every other line to the principal line, the common velocity, and width.
        The retrieved fluxes were dereddened through the A$_{Ks}$ values computed from local average spectrum. 

        \subsubsection{Uncertainties}
        
        The statistical errors, given by the fitting procedure, are small (few percents) and do not reflect the true uncertainties. 
        The main source of error is the continuum evaluation, which over- or underestimates the intensity of a given line. 
        This can affect the relative strength of lines and thus the quantities we wish to estimate.
        The line width and velocity are not much affected: thanks to the simultaneous fit, they are evaluated through many lines at the same time.
        
        To evaluate the effect on the line intensity, we considered for each spectral window the difference between the model and the continuum-subtracted spectrum.
        Then, we computed the interquartile range of this difference. 
        Half of the interquartile range provides the error estimate on the chosen continuum, thus on the intensity.
        
        We fit the flux that is related to intensity (I) and width ($\sigma$) through $F = I \cdot \sigma \cdot \sqrt{2\pi}$. 
        Therefore the final uncertainty on flux is $\sqrt{2\pi} \cdot ( I \cdot \Delta\sigma + \sigma \cdot \Delta I)$, where $ \Delta\sigma$ is the statistical uncertainty on the width and $\Delta I$ the uncertainty on intensity, estimated through the
interquartile range.

        \subsubsection{Fit of excitation diagrams}

        Given the dereddened line flux $f$ of a line, the column density of molecules in state $[v,j]$ is $N_{vj}=4\pi f g_{vj} /(\mathcal{A}_{vj}\Omega)$, where $g_{vj}$ is the degeneracy of the considered state and $\Omega$ is the aperture solid angle. 
        The excitation diagram shows $N_{vj}/g_{vj}$ versus the energy of the upper level.

        For thermalized populations, $N_{vj}$ is characterized by a Boltzmann distribution:
        \begin{equation}
        \frac{N_{vj}}{N_{tot}}=\frac{ g_{vj} e^{-E_{vj}/T_{e}} }{ \sum_{i} g_{i} e^{-E_{i}/T_{e}} }
        \label{Boltzmann}
        ,\end{equation}
        where $N_{tot}$ is the total column density, $E_{vj}$ is the upper level energy, and $T_{e}$ is the excitation temperature. 
        For an arbitrary excitation mechanism, a distinct excitation temperature might characterize each level. 
        The excitation temperature is representative of the kinetic temperature only in LTE.
        For more details see \cite{Goldsmith99}.
        
        We fit a thermal distribution to the observed points,   
        \begin{equation}
        \frac{N_{vj}/g_{vj}}{N_{13}/g_{13}} = a \cdot e^{-(E_{vj}-E_{13})/T_{e}}
        \label{fitED}
        ,\end{equation}          
        where $T_{e}$ is the excitation temperature and $a$ is the y-intercept in the log space. 
        The model can be fit using either all observed H$_{2}$ lines or  the ortho and  para lines independently. 
        The first case is equivalent to assuming that the ortho-para ratio (OPR) has the standard value of 3, which comes from spin statistics. 
        The second case allows testing whether the distribution is thermal, but with a non-standard OPR. 
        Each series of lines could then appear to be well fit by two parallel straight lines. In this case, the offset would correspond to the actual OPR.

        The uncertainties on the fit parameters of Eq.~\ref{fitED} have to be carefully considered. 
        Many sources of errors contribute to the final uncertainty, and their propagation is not straightforward. 
        Moreover, the two fitting parameters are correlated, and a classical analytical propagation is not applicable. 
        To overcome these difficulties, we considered the $\chi^{2}$ map in the ($a$,  $1/T_{e}$) plane. 
        This particular pair of parameters was adopted because it yields the most linear relation in the log space.
        The minimum on the $\chi^{2}$ map corresponds to the best estimates of $a$ and $1/T_{e}$ parameters. 
        The thermal model corresponding to these best estimates is traced on the excitation diagrams. 
 
        The associated uncertainties can be estimated through the $\chi^{2}$ maps. 
        To do so, we ran a Monte Carlo simulation of 10\,000 sets of column densities, normally distributed, according to the known uncertainties. 
        Then we fit each simulated set to obtain an estimate of $a$ and  $1/T_{e}$. 
        We calculated the $\chi^{2}$ value of this best-fit model when compared to the measured column densities. 
        With these 10\,000 $\chi^{2}$ values we traced the contours that enclose the 68\%, 95\%, and 99\% (respectively 1, 2, and 3-$\sigma$) of the smallest $\chi^{2}$ obtained.
        We calculated the final uncertainties on the fit parameters, considering the highest and lowest values of $a$ and $1/T_{e}$ that correspond to the 1-$\sigma$ contour. 
        
\subsection{Results of the excitation diagram analysis}
        
\begin{figure}
                \centering
                \includegraphics[width=9cm]{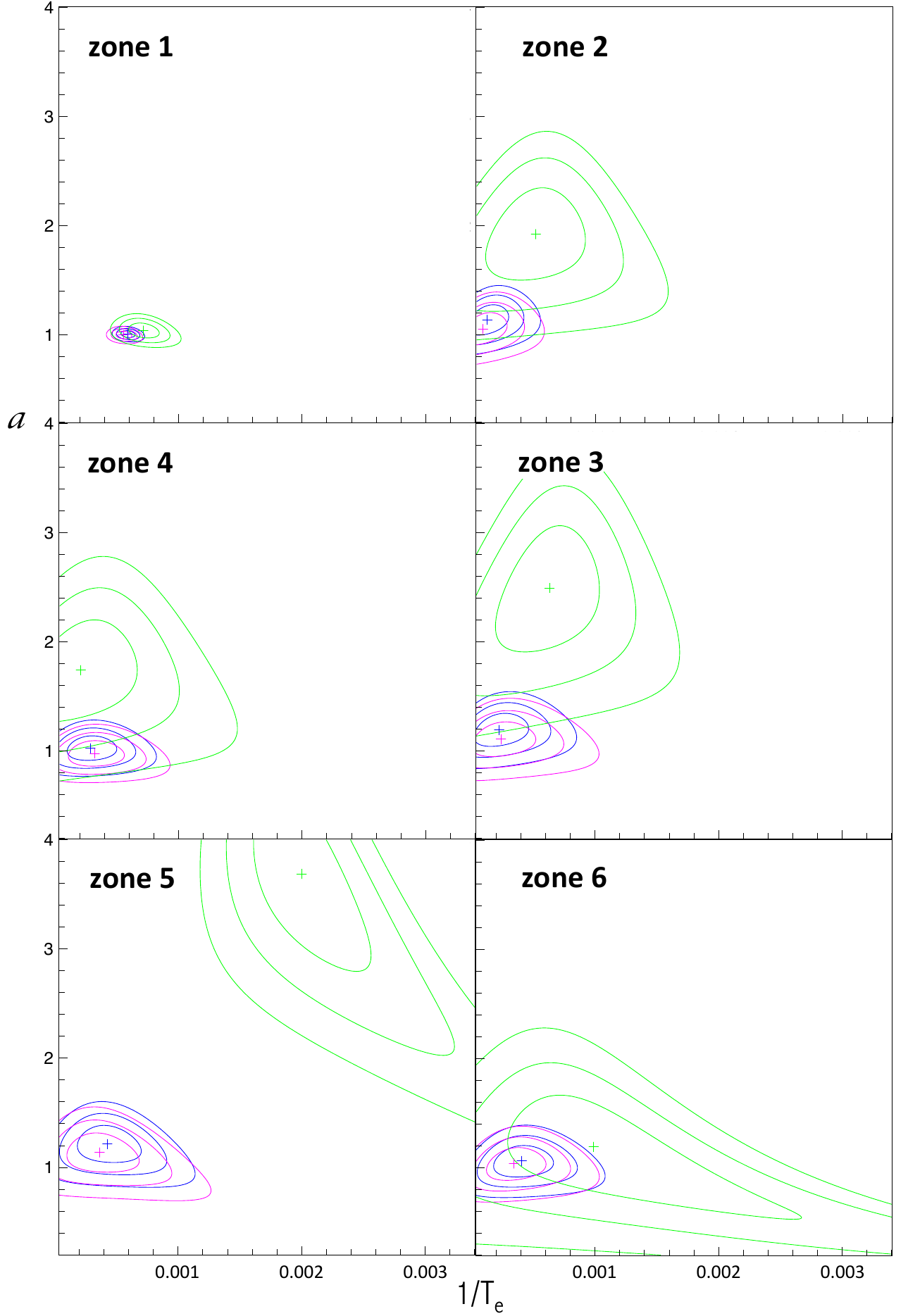}
                \caption{$\chi^{2}$ maps in the [ $1/T_{e}$, $a$]  plane for each zone when adjusting a thermal distribution to the data points. 
                The cross indicates the minimum of the $\chi^{2}$. 
                Contours indicate $\chi^{2}$ values corresponding to 1-, 2-, and 3-$\sigma$ for the all-lines (blue), ortho-lines (magenta), and para-lines (green) models.} 
                \label{res:chi2maps}
                \end{figure}

\begin{figure*}
                \centering
                \includegraphics[width=18.5cm]{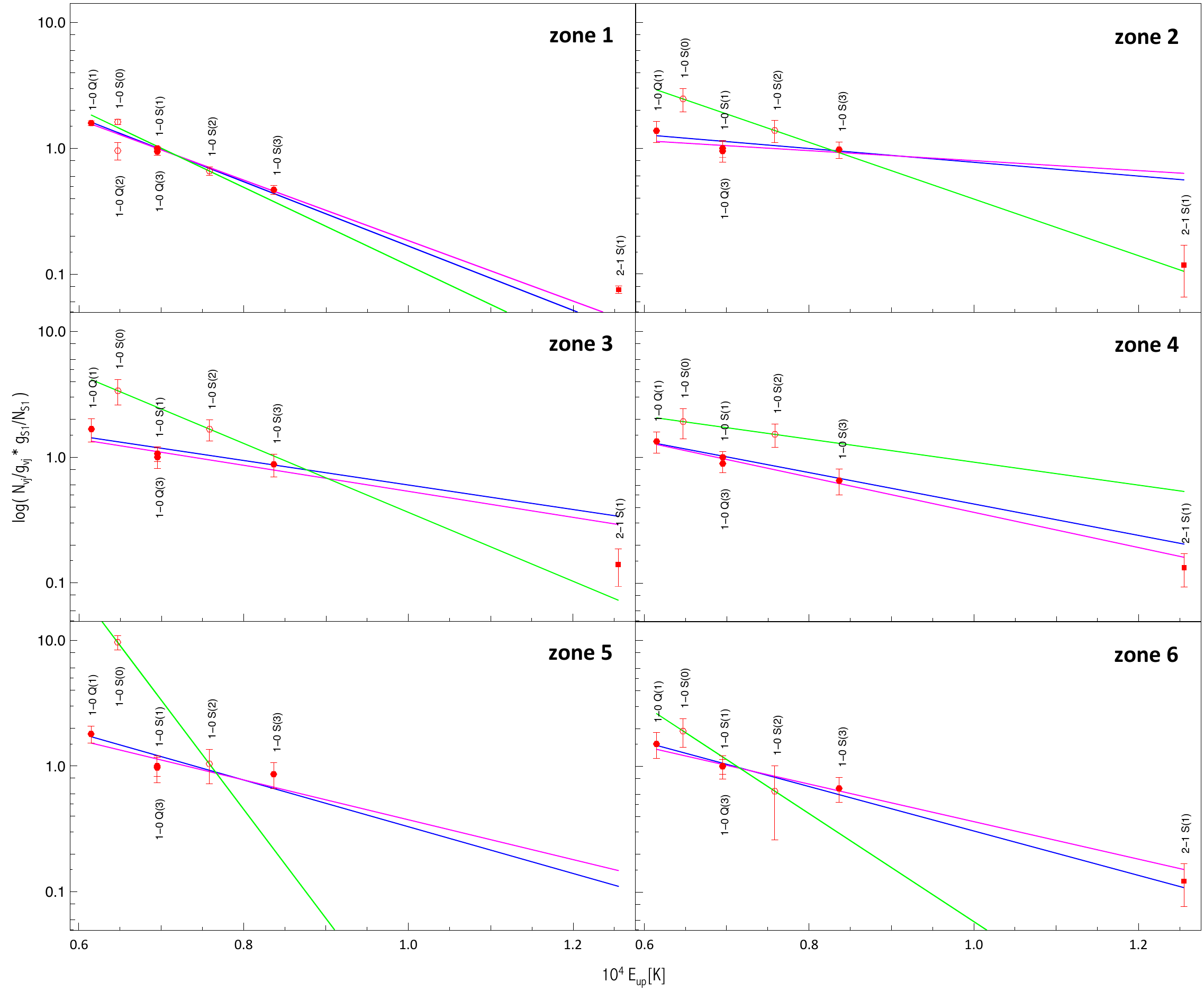}
                \caption{Excitation diagram (column density $log(\frac{N_{vj}/g_{vj}}{N_{13}/g_{13}})$ versus upper level energy expressed in K) for each zone.  
                Colored straight lines correspond to the best thermal model for all-lines (blue), ortho-lines (magenta), and para-lines (green).}
                \label{res:exc}
                \end{figure*}

        Figures~\ref{res:chi2maps} and \ref{res:exc} compare, for the six selected zones, the $\chi2$ maps and the excitation diagrams fit obtained for different sets of lines (all 1--0~lines, ortho or para lines alone), represented in different colors. 
        The objective of this comparison is to infer whether in a given zone the gas is thermalized. 
        Again only six zones are displayed, but the detailed information on the other zones is given in Appendix~\ref{A}.
        
        We considered that the population is thermalized when the all-lines, ortho-lines, and para-lines models are compatible within the uncertainties. 
        This translates into overlapping the three 1-$\sigma$ contours in the $\chi2$ space.  
         
                \begin{table*}[htb]
                \centering          
                \begin{tabular}{ c c | c c | c c | c c |}     
                \cline{3-8}   
                        &  & \multicolumn{2}{|c|}{all-lines model}   &   \multicolumn{2}{|c|}{ortho-lines model}  & \multicolumn{2}{|c|}{para-lines model}  \\[6pt]     
                \hline  
                \multicolumn{1}{ |c  }{zone} &  \multicolumn{1}{ |c | }{\scriptsize{1--0~S(1)/2--1 S(1)}} &
                        $T_{e}$ & $N_{tot}$  &
                        $T_{e}$ & $N_{tot}$  &
                        $T_{e}$ & $N_{tot}$  \\ [6pt]
                \hline          
                \multicolumn{1}{ |c  }{1} & \multicolumn{1}{ |c | }{9.8} & $1700_{-150}^{+150}$ & $27_{-7}^{+11}$ & $1800_{-150}^{+200}$ & $22_{-6}^{+10}$ &  $1400_{-150}^{+280}$ & $53_{-27}^{+66}$ \\ [6pt]
                \multicolumn{1}{ |c  }{6 }& \multicolumn{1}{ |c  |}{6}    & $2460_{-940}^{+3030}$ & $4_{-2}^{+11}$ & $2910_{-940}^{+5010}$ & $3_{-1}^{+8}$ &  $1010_{-940}^{+2470}$ & $>1$\\[6pt]
                        \hline               
                \end{tabular} 
                \vspace{0.5cm}
                \caption{1--0~S(1) to 2--1 S(1) ratio, excitation temperature (in K), and total column density (in 10$^{22}$m$^{-2}$) for zones where thermalization applies. 
                These values are estimated with a fit of all-lines, ortho-lines and para-lines (where possible) with a thermal distribution. }             
                \label{table:TN}      
                \end{table*}

        The all-lines model is always dominated by the ortho-lines contribution because the ortho lines are the most numerous. 
        The para-lines model is always slightly steeper, tilted toward higher values of $1/T_{e}$ and $a$ with respect to the other two models, but it remains compatible within 1-$\sigma$ for zones 1 and 6. 
        These zones correspond to regions in the CND. 
        
        In zone 5 the ortho and para models are significantly different by well above 3-$\sigma$. 
        Here, the para-lines model gives lower temperatures and higher values of $a$, with larger uncertainties than the ortho- and all-lines models. 
        The para-lines model exhibits a 5-$\sigma$ offset on the $\chi2$ map with respect to the ortho- and all-lines models. 
        For zones 3 the ortho- and para-lines models show a 2-$\sigma$ offset. 
        Even though for this zone the effect is less striking, both zones are most probably not thermalized.
        
        For the other zones (4 and 2) a tilt can be seen but not to a significant level. 
        These areas are located at the border of different specific regions, so that the observed behavior may be the result of an average between different regimes.

        Table $\ref{table:TN}$ gives our estimate of the excitation temperature for the zones that are compatible with thermal equilibrium.
                
\subsection{Discussion}
                        
        \subsubsection{Thermalized zones: the CND}
        
        For regions where thermalization applies, $T_{e}$ corresponds to the kinetic temperature of the gas, and the total column density of excited H$_{2}$ can be derived from the model parameters:
        \begin{equation}
        N_{tot}=\frac{N_{13}}{g_{13}} a \cdot e^{E_{13}/T_{e}}\sum_{vj}g_{vj} e^{-E_{vj}/T_{e}}
        .\end{equation}
        
        \noindent\textit{Excitation temperature and column density}\\ 
        
        The derived temperatures are quite high, ranging from $1\,700$~K to $3\,100$~K, but below the H$_{2}$ dissociation temperature ($\sim4\,000$~--~$5\,000$~K). 
        The values are very high with respect to previous measurements (400 K in \citealt{Genzel85}, 200~--~300~K in \citealt{Mills15} in the CND). 
        On the other hand, very high temperatures have previously been observed in other analyses of NIR H$_{2}$ lines, such as in the Crab nebula, where \cite{Loh11} found temperatures in the range of $2\,000$~--~$3\,000$~K using several H$_{2}$ lines in the range 2.02~--~2.33~$\mu$m (among others the 1--0~S(0), S(1), S(2) lines).
        
        Knowing the total column density, the gas mass can be roughly estimated as $N_{tot} D^{2} S  M_{H_{2}}/ M_{\odot}$, where $S$ in the solid angle covered by the region in steradians and $M_{H_{2}}$ the mass of a single H$_{2}$ molecule. 
        For instance, we considered the thermalized all-lines model of zone 1 across a surface of $\sim$~13~arcsec$^{2}$ and a distance to the Galactic Center $D = 8$~kpc. 
        The mass of excited H$_{2}$ contained in zone 1 is thus $7_{-3}^{+4}~10^{-3}$~M$_{\odot}$. 
        This is much lower than previous estimates.
        \cite{Christopher05} studied the HCN emission in the CND and obtained a typical mass per cloud of 2--3~10$^{4}$~M$_{\sun}$~and a total gas mass of 10$^{6}$~M$_{\sun}$. 

        In summary, the analysis of H$_{2}$ lines shows that the inferred temperatures are much higher and the masses apparently much lower than is estimated with observations of other molecules (CO, NH$_{3}$). 
        This is because the molecular hydrogen detected through those infrared lines is only the most excited one.
        In the determination of the column density, hot H$_{2}$ dominates the intensity, making the results very sensitive to the thermal structure of the cloud. 
        Only a small fraction of H$_{2}$ , but at high temperature,  is responsible for the observed column densities. 
        If the portion of the CND that is covered by the SPIFFI dataset (at least the equivalent of one extended cloud) had a mass of 10$^{4}$~M$_{\sun}$, at 400~K, it follows from Eq.~\ref{Boltzmann} that the flux of the 1--0~S(1) line would be almost 10 times weaker than observed.
        
        The emission from excited H$_{2}$ is likely to come from the UV illuminated skin of the clouds that form the CND. 
        Only the H$_{2}$ located in a thin shell at the surface of these clouds would be heated by either the UV field or by collisions, presumably from stellar winds or cloud-cloud collisions. 
        This interpretation is in agreement with the observed width map (Sect.~\ref{gasdyn}). 
        
        Simulation made with the Meudon code \citep{Le-Petit06} of the standard conditions of a PDR are coherent with this result (J. Le Bourlot, private communication). 
        In these simulations the considered parameters are a total optical depth A$_{V}=10$, a density of 10$^{3}$~cm$^{-3}$, 100 times the interstellar radiation field, and solar abundances. 
        The result is that almost all excited gas is located in the A$_{V} = 0.14$ layer and 30\% of the emitting gas is in the A$_{V} = 0.014$ layer, that is, in 1/700 of the volume. 
        
        The huge initial discrepancy with the results of \cite{Christopher05} is largely reduced, but still there is a large gap, which may indicate that a high percentage of the hot gas is not seen at all because of a high dust opacity.  
        Considering that the PDR conditions in the Sgr~A* region are more extreme than elsewhere \citep{Guesten89}, with 90\% of the ionizing flux coming from OB supergiants and Wolf-Rayet stars \citep{Martins07}, an additional factor could be applied.
        
        This result is consistent with previous results from \cite{Loh11} for the Crab nebula and \cite{Scoville82} for Orion, for example.
        \cite{Loh11} proposed the same interpretation: the hotter H$_{2}$ is only a small fraction of the total cloud mass, but this source dominates the intensity. 
        They found that the H$_{2}$ traced by NIR lines is 1/1000 of the total mass. 
        Similarly, in Orion \cite{Scoville82} found a H$_{2}$ mass around 10$^{-2}$~M$_{\sun}$, significantly lower than predicted by models. 
        Hence, on one hand, the cooler molecular hydrogen is missed because its too low temperature prevents its detection, and on the other hand, the huge dust opacity must block most of the emission of the hot component.\\
        
                \noindent\textit{Excitation mechanism} \\
        
        The two main mechanisms that can excite the molecular hydrogen are UV pumping and shocks. 
        The 1--0~S(1) to 2--1 S(1) ratio (\citealt{Shull78}), is a useful tool in principle for distinguishing between these two possibilities. 
                
        The high 1--0~S(1) to 2--1 S(1) ratio in zone 1 (Table~\ref{table:TN}) is consistent with a shock-excited gas. 
        Even though for dense PDRs, UV-excited H$_{2}$ can produce a ratio that resembles shock-excited gas \citep{Sternberg89}, for zone 1 our result seems fairly consistent with previous  conclusions of a shock-driven excitation. 
        \cite{Yusef-Zadeh01}, for instance, argued that the emission arises from shocks  deriving from random motions of clouds for a clumpy medium or of turbulence for a more homogeneous one.
        
        For zone 6 the thermalizatio can be produced by either shock or collisional fluorescence (i.e. collisionally excited in gas heated by UV radiation). 
        Here, the 1--0~S(1) to 2--1 S(1) ratio drops midway between 10 and 1. 
        \cite{Burton93} interpreted these values as indicative of collisional fluorescence.
        Moreover, since this zone shows a ratio significantly lower than 10, shock excitation seems to be ruled out, which leaves collisional fluorescence as the favored scenario. 
        This is in contrast with the opposite corner of the SPIFFI field, which is more probably dominated by collisional excitation. 
                
        \subsubsection{Nonthermalized zones: the central cavity}

        The population in the central cavity is almost certainly not thermalized. 
        No kinetic temperature and total column density can be determined straightforwardly. 
        Nevertheless, the fitting parameters are reported in Appendix~\ref{A} for completeness and to compare the results obtained for the distinct sets of lines. 
        These values have to be interpreted as indicators of a trend of temperature and column density, and they have to be considered cautiously. 
        
        Again, the observed H$_{2}$ only represents the most excited fraction. 
        However, the density in the central
cavity drops by one order of magnitude \citep[among others]{Jackson93,Davidson92} and the interpretation of the emitting gas as a thin layer at the border of clouds does not necessarily apply. 

        For instance, the 1--0~S(1) width map and the extinction map indicate that part of the emission comes from inside the northern arm cloud. 
        This might explain that the H$_{2}$ emission is detected
from inside the northern arm cloud, but not from inside the CND.
        In the CND clouds the emission remains confined at the surface because of the higher density, which allows an efficient self-shielding. 
        On the contrary, the clumpy environment of the northern arm cloud and the stronger UV field allow the radiation to penetrate deeper inside the cloud.
        
        Elsewhere in the central cavity, the gas might be more diffuse and distributed in larger volumes. 
        In this case, the observed intensities might represent a higher percentage of the total gas mass. \\
        
        \noindent\textit{Excitation mechanism}\\
        
        The excitation diagram shape, caused by nonthermalized gas, suggests that the excitation is provided by other sources than shock, such as UV or X-rays. 
                
        Indeed, the shape of the excitation diagram can be reproduced by a UV-pumped mechanism (E. Bron private communication). 
        However, this model fails to reproduce the high intensities we observe. 
        The central cavity environment is very complex, and several mechanisms may contribute to excite the gas. 
        Different interpretations probably apply to the northern arm cloud emission, the region south of GCIRS 7 and the minicavity (zone 5). 
        
        In particular, zone 5 is the region that corresponds to the strong flux enhancement of the para line 1--0~S(0) seen in its flux map (cf. Fig.~\ref{res:orthoMaps}, Sect.~\ref{orthomaps}). 
        This zone is particularly challenging to explain. 
        In this region all 1--0~lines align quite well on a straight line and only  S(0) strongly deviates in the excitation diagram. 
        We note that the other para line (1--0~S(2)) does not deviate, which excludes an OPR effect. 
        A possible explanation of this discrepancy might be found in the timescale of the H$_{2}$ formation and destruction process.
        The tentative explanation that we propose is that
        \begin{itemize}
        \item [a)] the central cavity has a higher density of UV photons and cosmic rays; 
        \item [b)] in an energetic region like this, H$_{2}$ molecules are more rapidly destroyed and therefore have a shorter mean lifetime, during which thermalization cannot fully occur; 
        \item [c)] if this holds true, we would mostly observe recently formed H$_{2}$ molecules that may form mainly as the para isomer  and  did not yet reach the equilibrium where OPR~=~3; 
        \item [d)] in addition, it is possible for molecular hydrogen to be formed in a state where peculiar energetic states are favored, such as the upper level of S(0), for instance. 
        \end{itemize}
        Of course this hypothesis requires a realistic modeling to be supported in a more precise way, and work has been started in this direction.
        Finally, a modeling involving UV pumping seems very promising to explain the minicavity emission. 
        A more detailed study of this region will be the object of a future research.

\section{Conclusions}
\label{concl}

We have studied the NIR H$_{2}$ emission in the central parsec of the Galaxy. Our analysis was divided into two main parts.

In the first part (Sect.~\ref{morph}) we studied the gas distribution and dynamics. 
Thanks to the new regularized 3D fitting, we were able to probe the morphology of H$_{2}$ in the central cavity. 
We provided a high-resolution map of the 1--0~S(1) line emission together with its velocity and width maps. 
The same was done for the other detectable lines. 
The flux maps are compatible with previous studies, and the velocity of the CND is consistent with its known dynamics. 
The new method allowed us to build an extinction map for the molecular gas as well.  
The shapes of the flux and of the width maps together suggest that there are two main components of the emission. 
In addition to the CND emission, a second component corresponds to the central cavity and it might be situated in the background, that is, behind the minispiral or in the central parsec itself. 

This ambiguity was clearly solved in the second part of our study (Sect.~\ref{prop}) through the multiline analysis. 
The comparison between the fluxes of each of the detectable lines allowed us to draw the excitation diagrams of the gas in several zones of the field. 
For the CND we found very high temperatures and very low masses compared to estimates obtained through CO observations. 
The reason is that the lines we observe trace only the very hot gas. 
This gas is a small fraction of the total amount, a thin layer at the surface of cooler clouds.
In addition, we had to invoke high dust opacities to explain the huge discrepancy.

The observation of the emission from the northern arm cloud confirms the close relationship between the CND and the minispiral. 
In the rest of the central cavity we observed a very high emission of S(0), especially in the zone close to the entrance of the 
minicavity. Here H$_{2}$ probably has a short mean lifetime because of the strong UV radiation. 
We tentatively proposed that the quick processes of destruction and formation of new H$_{2}$ molecules favors S(0) over other lines. 

This preliminary analysis shows that there are several different
areas in a volume as small as the zone inside the rim of the CND. Each feature shows some very peculiar conditions that can be attained only in an extreme environment such as the central parsec. 

Since we did not observe evidence of a motion of H$_{2}$ from the CND toward the central cavity, this molecular gas might be
thought to be produced locally, but the mechanism is still unclear. 
One possible interpretation, supported by simulations by \cite{Cuadra05}, is that the combined wind from the many mass-loosing stars in the central parsec provides the material and energy to build a plasma, as well the dust grains on which H$_{2}$ forms. If
this were found to be true, it
would explain our observations. 
On the other hand, the observed emission might arise from H$_{2}$ produced in denser regions and then brought into the central cavity by small-scale inward-motions that are able to escape detection in the velocity map. 
Additional studies are needed to conclude on one or the other hypothesis.
 

\begin{acknowledgements}
We would like to thank the infrared group at the Max Planck 
Institute for Extraterrestrial Physics (MPE), who built the SPIFFI instrument, carried out the observations, and provided the data.
We thank E. Bron, J. Le Bourlot and F. Le Petit for the simulations of H$_{2}$ in PDR and useful discussions. We are also grateful to N. Scoville for his comments and suggestions. This work is partially supported by the OPTICON project (EC FP7 grant agreement 312430) in ``Image reconstruction in optical interferometry'' WP4.
\end{acknowledgements}

 \newpage               
\bibliographystyle{aa} 
\bibliography{bibH2}{}

\begin{onecolumn}
\newpage

\begin{appendix} 

\section{Plots and tables for all sixteen zones}
\label{A}

                \begin{figure*}
                \centering
                \includegraphics[width=18cm]{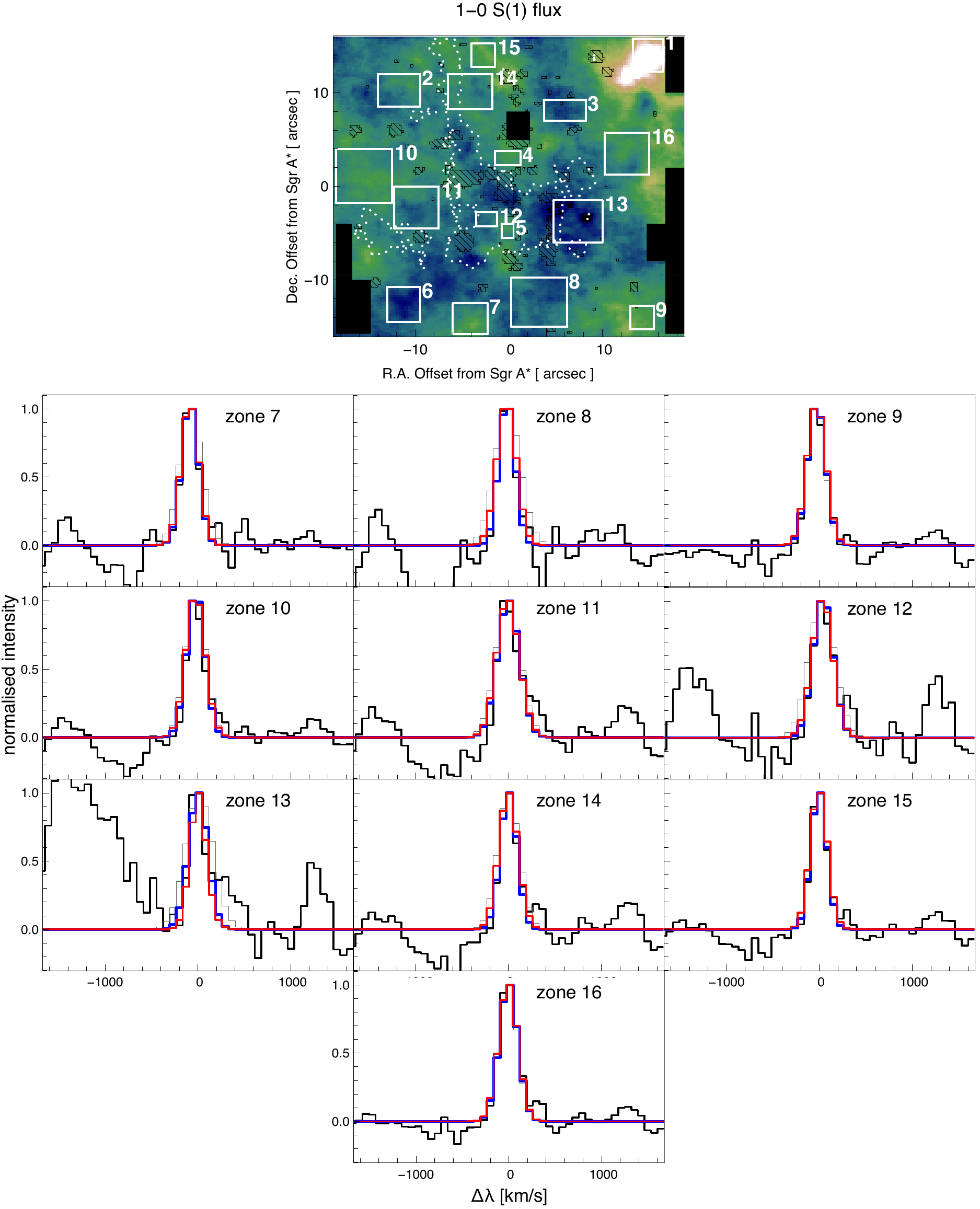}
                \caption{\textit{Top:} Zones where individual spectra of S(1) line have been extracted, superimposed on the
flux map. 
                \textit{Bottom:} for each zone we plot the normalized average spectrum (black) superimposed with the respective pixel by pixel 1D model (solid blue) and regularized 3D model (solid red). 
                The gray line reproduces another solution of the regularized 3D fitting. 
                Both solutions have similar values of $\chi2$ (1.15 and 1.49, respectively), but the red model fits the observations
much better.}
                \label{res:zone_all}
                \end{figure*}

                \begin{figure*}[htb]
                \centering
                \vspace{-0.2cm}
                \includegraphics[width=17.9cm]{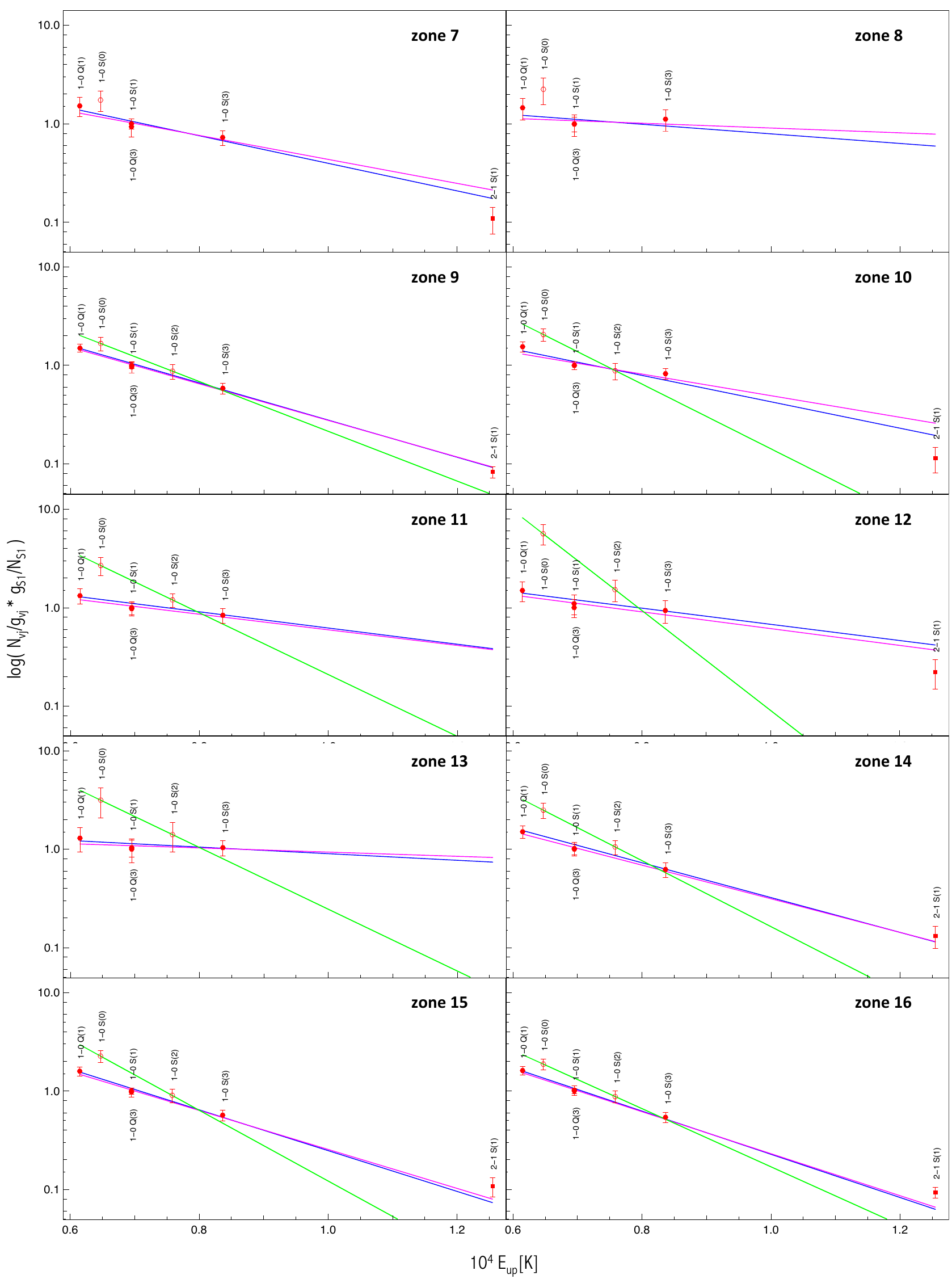}
                \vspace{-0.2cm}
                \caption{Excitation diagram (column density $log(\frac{N_{vj}/g_{vj}}{N_{13}/g_{13}})$ versus upper level energy expressed in K) for each zone.  
                Colored straight lines correspond to the best thermal model for all-lines (blue), ortho-lines (magenta), and para-lines (green) models.}
                \label{res:exc_all}
                \end{figure*}
                
                \begin{table*}[htb]
                \centering          
                \begin{tabular}{ c c | c c | c c | c c |}     
                \cline{3-8}   
                         &  & \multicolumn{2}{|c|}{all-lines model}   &   \multicolumn{2}{|c|}{ortho-lines model}  & \multicolumn{2}{|c|}{para-lines model}  \\[6pt]     
                \hline  
                \multicolumn{1}{ |c  }{zone} &  \multicolumn{1}{ |c | }{\scriptsize{1--0~S(1)/2--1 S(1)}} &
                        $T_{e}$ & $N_{tot}$  &
                        $T_{e}$ & $N_{tot}$  &
                        $T_{e}$ & $N_{tot}$  \\ [6pt]
                \hline          
                \multicolumn{1}{ |c  }{1} & \multicolumn{1}{ |c | }{9.8} & $1700_{-150}^{+150}$ & $27_{-7}^{+11}$ & $1800_{-150}^{+200}$ & $22_{-6}^{+10}$ &  $1400_{-150}^{+280}$ & $53_{-27}^{+66}$ \\ [6pt]
                \multicolumn{1}{ |c  }{6 }& \multicolumn{1}{ |c  |}{6}    & $2460_{-940}^{+3030}$ & $4_{-2}^{+11}$ & $2910_{-940}^{+5010}$ & $3_{-1}^{+8}$ &  $1010_{-940}^{+2470}$ & $>1$\\[6pt]
                \multicolumn{1}{ |c  }{7} & \multicolumn{1}{ |c  |}{6.8} & $3100_{-1220}^{+4400}$ & $7_{-1}^{+13}$ & $3560_{-1220}^{+7400}$ & $6_{-4}^{+10}$ & - & - \\ [6pt]
                \multicolumn{1}{ |c  }{ 9} & \multicolumn{1}{ |c | }{9}    & $2300_{-450}^{+730}$ & $6_{-2}^{+5}$ & $2340_{-450}^{+830}$ & $6_{-2}^{+5}$ &  $1720_{-450}^{+1940}$ & $15_{-11}^{+90}$ \\ [6pt]
                \hline
                \end{tabular} 
                \vspace{0.5cm}
                \caption{1--0~S(1) to 2--1 S(1) ratio, excitation temperature (in K), and total column density (in 10$^{22}\cdot$m$^{-2}$) for each zone where thermalization applies. 
                These values are estimated with a fit of all-lines, ortho-lines, and para-lines models (where possible) with a thermal distribution.}
                \label{NH2table_alltherm}      
                \end{table*}

                \begin{table*}[htb]
                \centering          
                \begin{tabular}{ cc | c c | c c | c c |}     
                \cline{3-8}   
                        &  &\multicolumn{2}{|c|}{all-lines model}   &   \multicolumn{2}{|c|}{ortho-lines model}  & \multicolumn{2}{|c|}{para-lines model}  \\[6pt]     
                \hline  
                \multicolumn{1}{ |c  }{zone} &  \multicolumn{1}{ |c | }{\scriptsize{1--0~S(1)/2--1 S(1)}} & 
                        $T_{e}$ & $N_{tot}$   &
                        $T_{e}$ & $N_{tot}$   &
                        $T_{e}$ & $N_{tot}$   \\ [6pt]
                \hline               
                \multicolumn{1}{ |c  }{2} &  \multicolumn{1}{ |c | }{6.3} & $>3390$ & $<5$ & $>6440$ & $<10$ &  $<8910$ & $<168$ \\ [6pt]
                \multicolumn{1}{ |c  }{3} &  \multicolumn{1}{ |c | }{5.3} & $4460_{-2260}^{+43070}$ & $2_{-16}^{+3}$ & $4320_{-2260}^{+138250}$ & $2_{-30}^{+5}$ &  $1580_{-2260}^{+3700}$ & $21_{-18}^{+276}$ \\ [6pt]
                \multicolumn{1}{ |c  }{4} &   \multicolumn{1}{ |c | }{5.6} &$>3333$ & $<13$ & $>1630$ & $<17$ &  $>3280$ & $<58$ \\ [6pt]
                \multicolumn{1}{ |c  }{5} &  \multicolumn{1}{ |c | }{-} & $2340_{-910}^{+2940}$ & $5_{-3}^{+21}$ & $2740_{-910}^{+6760}$ & $4_{-0}^{+18}$ &  $<620$ & $>8640$ \\ [6pt]
                \multicolumn{1}{ |c  }{8} &  \multicolumn{1}{ |c | }{-} & $>2130$ & $<6$ & $>13000$ & $<19$ & - & - \\ [6pt]
                \multicolumn{1}{ |c  }{10} &  \multicolumn{1}{ |c | }{6.5} & $3240_{-1010}^{+2460}$ & $4_{-1}^{+3}$ & $3960_{-1010}^{+4430}$ & $3_{-1}^{+2}$ &  $1320_{-1010}^{+940}$ & $38_{-32}^{+341}$ \\ [6pt]
                \multicolumn{1}{ |c  }{11} &  \multicolumn{1}{ |c | }{-} & $>2850$ & $<5$ & $3050$ & $<5$ &  $<3030$ & $>4$ \\ [6pt]
                \multicolumn{1}{ |c  }{12} &  \multicolumn{1}{ |c | }{3.3} & $>2230$ & $<7$ & $>2040$ & $<9$ &  $<1440$ & $>25$ \\ [6pt]
                \multicolumn{1}{ |c  }{13} &  \multicolumn{1}{ |c | }{-} & $>3960$ & $<19$ & $>30000$ & $<32$ & $<140000$ & $>12$ \\ [6pt]
                \multicolumn{1}{ |c  }{ 14} &   \multicolumn{1}{ |c | }{5.6} &$2500_{-670}^{+1250}$ & $3_{-1}^{+5}$ & $2550_{-670}^{+1650}$ & $3_{-1}^{+5}$ &  $1300_{-670}^{+1040}$ & $32_{-28}^{+265}$ \\ [6pt]
                \multicolumn{1}{ |c  }{15} &  \multicolumn{1}{ |c | }{6.8} & $2070_{-390}^{+620}$ & $8_{-4}^{+8}$ & $2190_{-390}^{+780}$ & $7_{-3}^{+8}$ &  $1210_{-390}^{+620}$ & $71_{-59}^{+414}$ \\ [6pt]
                \multicolumn{1}{ |c  }{16} &  \multicolumn{1}{ |c | }{7.9} & $1980_{-340}^{+480}$ & $7_{-3}^{+7}$ & $2040_{-340}^{+560}$ & $6_{-3}^{+6}$ &  $1470_{-340}^{+870}$ & $22_{-17}^{+94}$ \\ [6pt]
                \hline               
                \end{tabular} 
                \vspace{0.5cm}
                \caption{1--0~S(1) to 2--1 S(1) ratio and best-fit parameters obtained when considering a thermal distribution on all-, ortho-, and para-lines models for each zone. 
                These parameters values are valid under the assumption of thermalized populations, which does not apply to these regions. 
                They therefore have to be considered with caution as indicators of a trend of temperature and column density. 
                Apparent $T_{e}$ is in K, apparent $N_{tot}$ is in 10$^{22}\cdot$m$^{-2}$.}             
                \label{NH2table_allnottherm}      
                \end{table*}

                \begin{figure*}
                \centering
                \hspace{-1cm}\includegraphics[width=10cm]{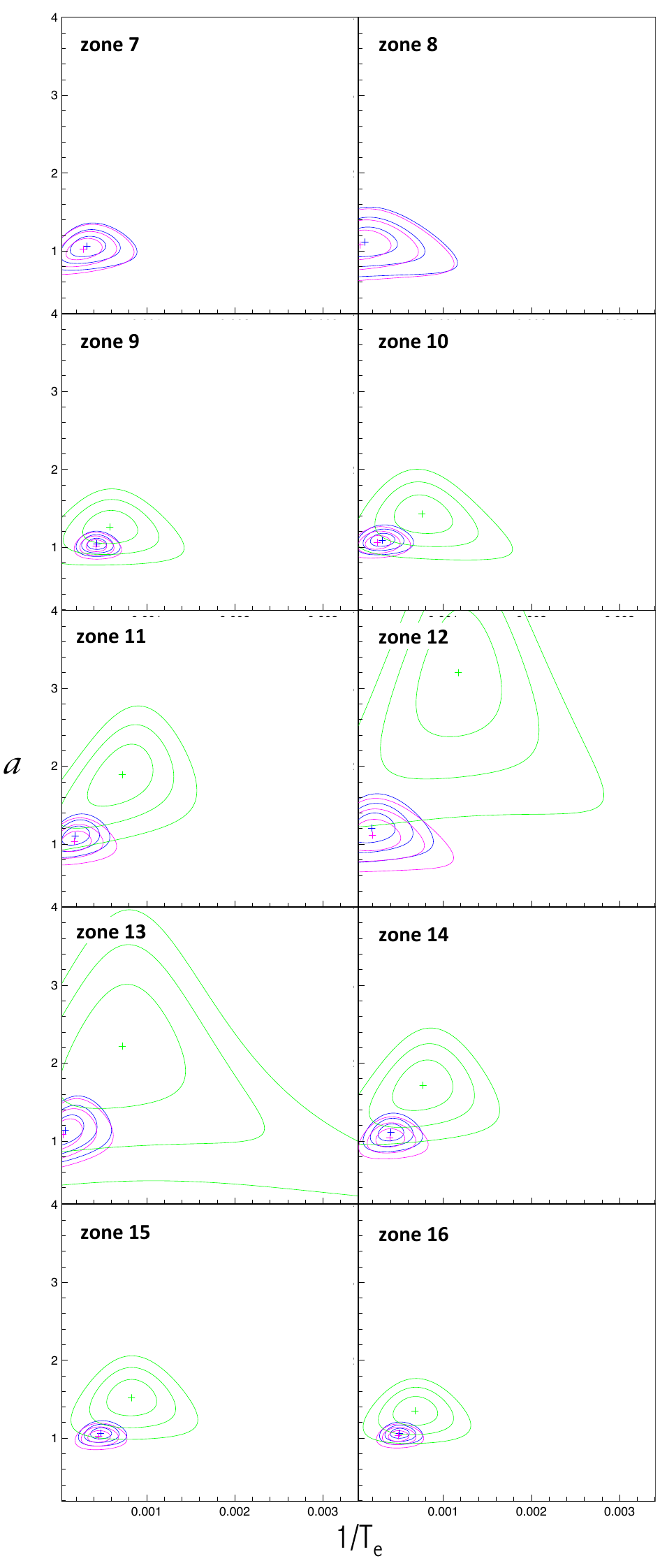}
                \caption{$\chi^{2}$ maps for each zone. 
                The cross indicates the minimum of the $\chi^{2}$. 
                Contours indicates the $\chi^{2}$ values corresponding to 1-, 2-, and 3-$\sigma$. 
                On the same map the all-lines (blue), ortho-lines (magenta), and para-lines (green) models are superimposed.}
                \label{res:chi2maps_all}
                \end{figure*}

\newpage

\end{appendix} 
\end{onecolumn}


\end{document}